\documentclass[conference]{IEEEtran}
\IEEEoverridecommandlockouts
\usepackage{cite}
\usepackage{caption}
\usepackage{subcaption}
\usepackage{float}
\usepackage{multirow}

\usepackage{caption}
\usepackage{subcaption}
\usepackage{pifont}
\usepackage{makecell}

\newcommand{\cmark}{\ding{51}}%

\newcommand{\GL}{\emph{EASE}}
\newcommand{\GLSS}{\emph{EASE's}}
\newcommand{\QP}{\emph{PartitioningQualityPredictor}}
\newcommand{\PrP}{\emph{ProcessingTimePredictor}}
\newcommand{\PaP}{\emph{PartitioningTimePredictor}}
\newcommand{\PS}{\emph{PartitionerSelector}}

\newcommand{\SPS}{$S_{\mathit{PS}}$}
\newcommand{\SPSS}{$S_{\mathit{PS}}$'s}

\newcommand{\SR}{$S_{\mathit{R}}$}

\newcommand{\SSRF}{$S_{\mathit{SRF}}$}

\usepackage{amsmath,amssymb,amsfonts}
\usepackage{algorithmic}
\usepackage{graphicx}
\usepackage{textcomp}
\usepackage{xcolor}
\def\BibTeX{{\rm B\kern-.05em{\sc i\kern-.025em b}\kern-.08em
    T\kern-.1667em\lower.7ex\hbox{E}\kern-.125emX}}
\begin{document}

\title{Partitioner Selection with EASE to Optimize Distributed Graph Processing}

\author{
    \IEEEauthorblockN{Nikolai Merkel\IEEEauthorrefmark{1}, Ruben Mayer\IEEEauthorrefmark{1}, Tawkir Ahmed Fakir\IEEEauthorrefmark{1} and Hans-Arno Jacobsen\IEEEauthorrefmark{2}}
    \IEEEauthorblockA{\IEEEauthorrefmark{1}Technical University of Munich, \IEEEauthorrefmark{2}University of Toronto\\
\{nikolai.merkel, ruben.mayer, tawkir.fakir\}@tum.de}
\IEEEauthorblockA{jacobsen@eecg.toronto.edu}}

\maketitle

\makeatletter
\def\ps@IEEEtitlepagestyle{
  \def\@oddfoot{\mycopyrightnotice}
  \def\@evenfoot{}
}
\def\mycopyrightnotice{
  {\footnotesize
  \begin{minipage}{\textwidth}
  \centering
(c) 2023 IEEE. Personal use of this material is permitted. Permission from IEEE must be obtained for all other uses, in any current
  or future media, including reprinting/republishing this material for advertising or promotional purposes, creating new collective works,
  for resale or redistribution to servers or lists, or reuse of any copyrighted component of this work in other works.
  The definitive version is published in Proceedings of the 2023 IEEE 39th International Conference on Data Engineering (ICDE '23).
  
  \end{minipage}
  }
}

\begin{abstract}
For distributed graph processing on massive graphs, a graph is partitioned into multiple equally-sized parts which are distributed among machines in a compute cluster. In the last decade, many partitioning algorithms have been developed which differ from each other with respect to the partitioning quality, the run-time of the partitioning and the type of graph for which they work best. The plethora of graph partitioning algorithms makes it a challenging task to select a partitioner for a given scenario. Different studies exist that provide \textit{qualitative} insights into the characteristics of graph partitioning algorithms that support a selection. However, in order to enable automatic selection, a \emph{quantitative} prediction of the partitioning quality, the partitioning run-time and the run-time of subsequent graph processing jobs is needed. In this paper, we propose a machine learning-based approach to provide such a quantitative prediction for different types of edge partitioning algorithms and graph processing workloads. We show that training based on generated graphs achieves high accuracy, which can be further improved when using real-world data. Based on the predictions, the automatic selection reduces the end-to-end run-time on average by 11.1\% compared to a random selection, by 17.4\% compared to selecting the partitioner that yields the lowest cut size, and by 29.1\% compared to the worst strategy, respectively. Furthermore, in 35.7\% of the cases, the best strategy was selected. 
\end{abstract}

\begin{IEEEkeywords}
graph partitioning, automatic partitioner selection, distributed graph processing, machine learning
\end{IEEEkeywords}

\section{Introduction}
\label{sec:indroduction}
In the last decade, many distributed graph processing systems and databases have emerged to process large graphs with billions of edges, such as Pregel \cite{pregel}, Giraph \cite{giraph}, PowerGraph \cite{powergraph}, PowerLyra \cite{powerlyra}, GraphX \cite{graphx} and Neo4j \cite{neo4j}. In order to enable distributed graph processing, a graph must be partitioned into multiple equally sized parts which are distributed among multiple machines in a compute cluster \cite{hdrf,dne,spinner}. Each machine performs computations on its partition and communication between the machines takes place via the network. The amount of communication between the machines is influenced by the quality of the partitioning and affects the performance of the distributed graph processing \cite{bourse, spinner, hep, survey.4.sigmod.2019}.

Many graph partitioning algorithms have been developed and can be categorized as follows: \emph{Streaming algorithms} \cite{fennel,stanton,CuSP,dbh,powerlyra,grid,graphx,boost,restreaming,adwise,twophasestreaming,2psl,windowvertexpartitioning} stream the graph, e.g., as an edge list and, dependent on the cut model, assign edges or vertices to partitions one after the other on the fly. \emph{In-memory algorithms} \cite{metis, ne, dne, sheep, schulz, spinner} load the entire graph into memory for partitioning. \emph{Hybrid algorithms} \cite{hep} partition one part of the graph in-memory and the remaining part in a streaming fashion. The partitioning algorithms differ from each other with respect to the partitioning quality, the run-time of the partitioning and the type of graph for which they work best \cite{survey.1.vldb.2017, survey.2.vldb.2018, survey.3.vldb.2018, survey.4.sigmod.2019, hep}.

The plethora of graph partitioning algorithms makes it a challenging task to select a partitioner for a given scenario.
It is hard to tell which partitioner will lead to the best partitioning quality on a given graph, or how much will be the quality difference between two partitioners. It is even harder to select the partitioning algorithm that leads to the lowest end-to-end run-time of both partitioning run-time and subsequent graph processing run-time. However, the choice of partitioning algorithm matters a lot. As we show in our paper, the best choice can lead to a reduction of end-to-end run-time of up to 71\% compared to the worst choice.

Existing studies \cite{survey.1.vldb.2017, survey.2.vldb.2018, survey.3.vldb.2018, survey.4.sigmod.2019} on graph partitioning algorithms provide interesting qualitative insights into the characteristics of these algorithms. For example, it was shown that there is not a single partitioner that works best for all graph processing workloads, that partitioning run-time is not always negligible and that the degree distribution of a graph can influence that partitioning result. However, the studies are not sufficient to automatically select a graph partitioner which minimizes the end-to-end time for a given graph and graph processing workload. In order to automatically choose among multiple partitioners, a \emph{quantitative} prediction of the partitioning run-time and the graph processing run-time is needed. For example, if the end-to-end time should be minimized, the sum of the predicted partitioning run-time and the graph processing run-time matters. 

We propose a machine learning-based approach to provide such a prediction and thereby enable automatic partitioner selection to either minimize the graph processing or end-to-end processing run-time. To this end, we profile different types of partitioners on a wide range of graphs and measure the partitioning run-time, partitioning quality metrics and properties of the graph. Then, we execute different graph processing algorithms on the partitioned graphs to measure the graph processing run-time. Finally, we use the profiled data to build machine learning models to predict the partitioning run-time and the partitioning quality metrics on an arbitrary graph for different partitioners. In addition, we train a model to predict the run-time of different graph processing algorithms based on the predicted partitioning quality metrics and graph properties. While there are different formalizations of graph partitioning, we focus on the edge partitioning problem in this work, as it is commonly used in many distributed graph processing frameworks \cite{graphx,powergraph,DistGNN} and subject to vibrant research \cite{hep, ne,dne,hdrf}.

A major challenge in using machine learning for the said prediction tasks is the need for training data which covers a wide spectrum of possible graphs (with different properties). It can be difficult to acquire enough representative real-world graphs for training for the following reasons. First, the number of available real-world graphs can be limited. Second, only known graphs can be used for the training, but new graphs with different properties may appear in future workloads. Third, the graphs may need to be downloaded from different sources in different formats which is a time-intensive and brittle process. An additional challenge is the selection of features for the machine learning models and how to fine tune the models. 

In our work, we make the following contributions:

\begin{enumerate}
	\item We propose \GL, a machine learning-based system to predict the partitioning and processing run-time for a given graph, graph partitioner and graph processing algorithm. We show that these predictions enable automatic edge partitioner selection which reduces the end-to-end run-time on average by 11.1\% compared to a random selection, by 17.4\% compared to selecting the partitioner that yields the lowest cut size, and by 29.1\% compared to the worst strategy, respectively. Furthermore, in 35.7\% of the cases, the best strategy was selected. Our approach is extensible so that new partitioners and graph processing algorithms can be incorporated without needing to re-train the entire model.
	\item We show how to tackle the challenge of acquiring a large variety of graphs for training by using a graph generator. This makes our approach easy to apply and reproduce, as we provide the graph generator along with its settings, so that the model can be trained without being dependent on external graph datasets. 
	\item We describe how the prediction accuracy of the machine learning model can be further improved by enriching the synthetic graphs with real-world graphs, if available. This enables the user to include real-world graphs in order to further increase the performance of the machine learning model for specific graph types. With enrichment, we decrease the prediction error for the replication factor by a factor of up to $3.2\times$ which leads to a reduction of end-to-end time of 5\% for the enriched type of graph.   
\end{enumerate}

The rest of the paper is organized as follows. In Section~\ref{sec:background}, we introduce the edge partitioning problem, partitioning quality metrics and common graph properties. In Section~\ref{sec:background:selection}, we show the importance of partitioning algorithm selection for graph processing performance. In Section~\ref{sec:approach}, we introduce our machine learning based approach for partitioning quality, partitioning run-time and processing run-time prediction. In Section~\ref{sec:evaluation}, we evaluate our approach. In Section~\ref{sec:relatedwork}, we discuss related work. Finally, we conclude in Section~\ref{sec:conclusion}. 
\section{Background}
\label{sec:background}
Let $G = (V, E)$ be a graph consisting of a set of vertices $V$ and a set of edges $E \subseteq V \times V$. The goal of edge partitioning is to cut G into $k$ partitions. In edge partitioning \cite{bourse}, the edges are divided into $k$ pairwise disjoint partitions $P = \{p_1, \dots, p_k\}$ with $\cup_{i=1}^{k}p_{i} = E$. Each partition $p_i$ covers a set of vertices $V(p_i) = \{v \in V | \exists u \in V :  (u, v) \in p_i \vee (v, u) \in p_i \}$. The vertices covered by a partition can be further categorized into source and destination vertices defined as $V_{\mathit{src}}(p_i) = \{u \in V | \exists v \in V : (u,v) \in p_i\}$ and $V_{\mathit{dst}}(p_i)= \{v \in V | \exists u \in V : (u,v) \in p_i\}$, respectively. Vertices which are covered by multiple partitions are cut and need to be replicated. The goal of edge partitioning \cite{ne} is to minimize the number of replicated vertices with the constraint that the edge partitions are $\alpha$-balanced: $\forall p_i \in P: |p_i| \leq \alpha \cdot \frac{|E|}{k}$. 

\subsection{Partitioning Quality Metrics}
\label{sec:background:partitioningqualitymetrics}
Different partitioning quality metrics exist. First, the \emph{replication factor} which is defined as $RF(P)=\frac{1}{|V|}\sum_{i \in [k]}|V(p_i)|$. It represents the average number of partitions a vertex spans. The replication factor is closely related to the communication cost \cite{bourse, survey.1.vldb.2017}. Second, different \emph{balancing metrics} that measure how balanced a partitioning is. In distributed graph processing, a balanced partitioning is important in order to avoid overloading machines that become stragglers \cite{survey.1.vldb.2017}. Often, the \emph{edge balance} is considered in vertex-cut partitioning (e.g., \cite{dne,bourse,hdrf,dbh,twophasestreaming,hep,adwise}). In addition, the \emph{vertex balance} can influence the processing performance \cite{hep,dne} which we also show in Section \ref{sec:background:selection}. For directed graphs, one can further differentiate between \emph{source vertex balance} and \emph{destination vertex balance} if the computation or communication load is dependent on the number of source or destination vertices per partition.  

Let $M = \{m_1, \dots, m_k\}$ be a set containing natural numbers. We define $max(M)$ as the largest element $x \in M$ s.t. $\forall y \in M: y \leq x$, and $avg(M)=\frac{1}{|M|}\sum_{i \in [k]} m_i$ as the average of the set. Now, we define four balancing metrics. 

\begin{enumerate}
    \item \emph{Edge balance:} $B_{\mathit{edge}}(P)=\frac{\mathit{max}({|p_1|, \dots, |p_k|})}{\mathit{avg}({|p_1|, \dots, |p_k|})}$
    \item \emph{Vertex balance:} $B_{v}(P)=\frac{\mathit{max}({|V(p_1)|, \dots, |V(p_k)|})}{\mathit{avg}({|V(p_1)|, \dots, |V(p_k)|}))}$
    \item \emph{Source balance:} $B_{\mathit{src}}(P)=\frac{\mathit{max}({|V_{\mathit{src}}(p_1)|, \dots, |V_{\mathit{src}}(p_k)|})}{\mathit{avg}({|V_{\mathit{src}}(p_1)|, \dots, |V_{\mathit{src}}(p_k)|}))}$
    \item \emph{Destination balance:} \newline $B_{\mathit{dst}}(P)=\frac{\mathit{max}({|V_{\mathit{dst}}(p_1)|, \dots, |V_{\mathit{dst}}(p_k)|})}{\mathit{avg}({|V_{\mathit{dst}}(p_1)|, \dots, |V_{\mathit{dst}}(p_k)|}))}$
    \end{enumerate}
    In this work, we build machine learning models to predict these five partitioning quality metrics (i.e., replication factor and the four balancing metrics) as a basis for graph processing run-time prediction. 

    \subsection{Graph Properties}
    \label{sec:background:graphproperties}
    There are different properties that characterize a graph. The following graph properties are used as features for our machine learning models. In Section \ref{sec:machine_learning_features}, a detailed discussion is provided why these graph properties are promising feature candidates for our prediction tasks.

    \begin{enumerate}
      \item \emph{Density:} The density is defined as $\mathit{dens}(G) = \frac{|E|}{|V|\cdot(|V| - 1)}$ and describes how many edges are contained in a graph compared to all possible edges which could be created with the given vertices \cite{fortunato}.
      \item \emph{Average degree:} The average degree is defined as \allowbreak $\mathit{deg}(G) = \frac{2\cdot|E|}{|V|}$. 
      \item \emph{Average number of triangles:} A triangle is a complete subgraph $T = \{ V_T, E_T\}$ of an undirected graph $G = (V, E)$ with exactly three vertices \cite{brandes2005}. The number of triangles of a vertex is defined as $\mathit{t}(v) = |\{T | v \in V_T\}|$. The average number of triangles is defined as $\mathit{t}(G) = \frac{1}{|V|}\sum_{v \in V} \mathit{t}(v)$. 
      \item \emph{Average local clustering coefficient (LCC):} The local clustering coefficient \cite{watts, newman, brandes2005} of a vertex $v$ is defined as \\ $\mathit{c}(v) = \frac{\mathit{t}(v)}{0.5\cdot\mathit{deg}(v)\cdot(\mathit{deg}(v)-1)}$, with $\mathit{deg}(v)$ being the degree of $v$ and $t(v)$ the number of triangles (see above). The average local clustering coefficient is defined as $\mathit{C}(G) = \frac{1}{|V|}\sum_{v \in V} \mathit{c}(v)$.
      \item \emph{Skewness:} For all metrics which are calculated for each vertex $v$ in the graph (e.g., degree), also, the skewness of the values can be described with Pearson's first skewness coefficient $\mathit{skew}(\mathit{values}) = \frac{\mathit{mean}(\mathit{values})-\mathit{mode}(\mathit{values})}{\sigma}$, with $\sigma$ being the standard deviation of the values.
    \end{enumerate}

\section{Graph Partitioner Selection}
\label{sec:background:selection}
As mentioned in Section \ref{sec:indroduction}, partitioning algorithms differ a lot from each other. The properties of graph partitioning algorithms pose a complex trade-off when selecting a partitioner. While a better partitioning quality in many cases leads to faster distributed graph processing, it may also be more expensive in terms of partitioning run-time to achieve it.

We showcase this in the following experiments for two graph processing algorithms: PageRank~\cite{pagerank} and Label Propagation~\cite{lpa}. We choose PageRank as a communication-bound algorithm which is sensitive to the replication factor and Label Propagation as a computation-bound algorithm which is sensitive to load balancing.

\subsection{PageRank}
\begin{figure}[t]
\centering
  \begin{subfigure}[b]{0.32\linewidth}
      \centering
      \includegraphics[width=\linewidth]{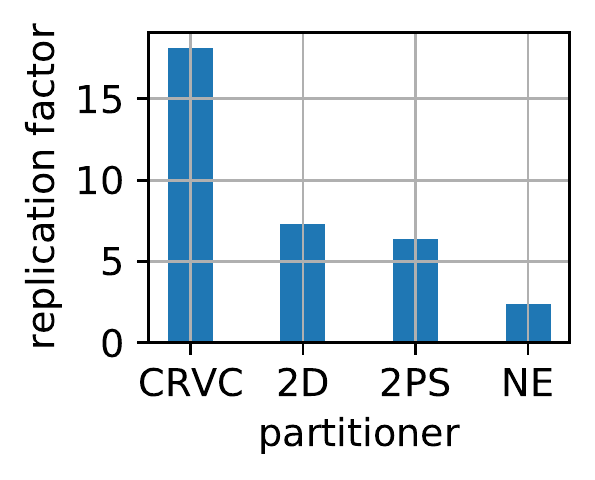}
      \vspace{-6mm}
      \caption{Replication factor (FR).}
      \label{fig:back:process-vs-replication_friendster-1}
  \end{subfigure}
  \hfill
\begin{subfigure}[b]{0.32\linewidth}
    \centering
    \includegraphics[width=\linewidth]{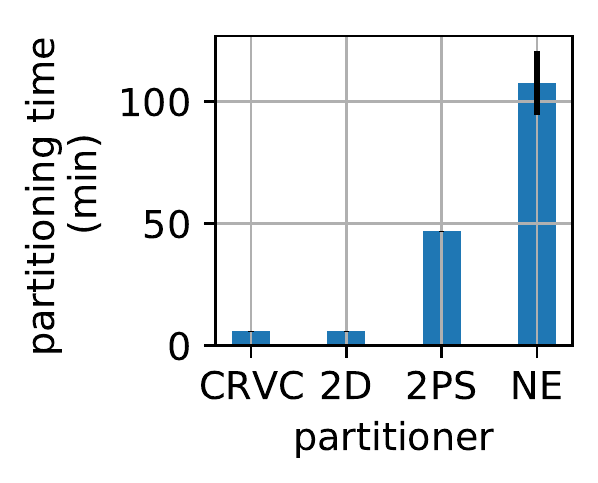}
    \vspace{-6mm}
    \caption{Partitioning \\  run-time (FR).}
    \label{fig:back:process-vs-replication_friendster-3}
\end{subfigure} 
\begin{subfigure}[b]{0.32\linewidth}
    \centering
    \includegraphics[width=\linewidth]{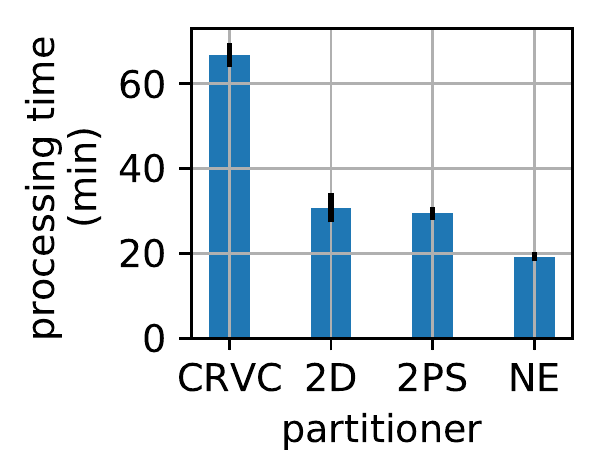}
    \vspace{-6mm}
    \caption{PageRank run-time (FR).}
    \label{fig:back:process-vs-replication_friendster-2}
\end{subfigure}
\begin{subfigure}[b]{0.32\linewidth}
  \centering
  \includegraphics[width=\linewidth]{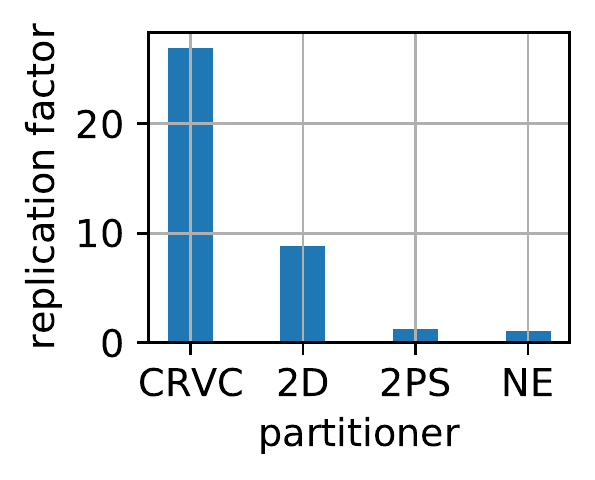}
  \vspace{-6mm}
  \caption{Replication factor (SK).}
  \label{fig:back:process-vs-replication_sk-2005-1}
\end{subfigure}
\hfill
\begin{subfigure}[b]{0.32\linewidth}
\centering
\includegraphics[width=\linewidth]{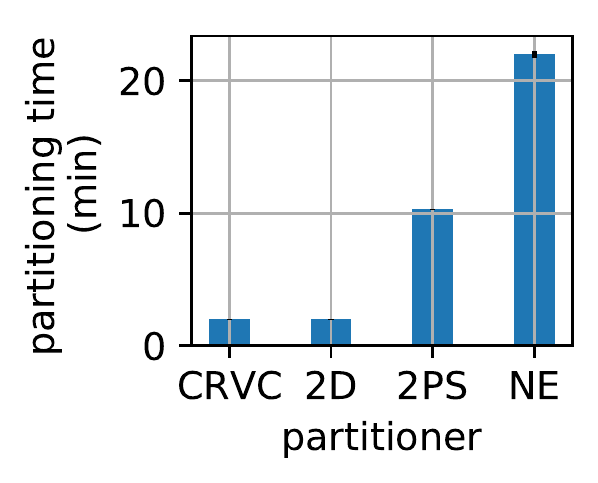}
\vspace{-6mm}
\caption{Partitioning \\ run-time (SK).}
\label{fig:back:process-vs-replication_sk-2005-3}
\end{subfigure} 
\begin{subfigure}[b]{0.32\linewidth}
\centering
\includegraphics[width=\linewidth]{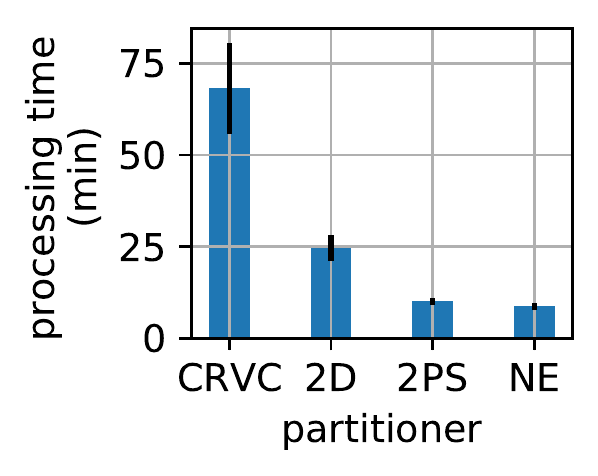}
\vspace{-6mm}
\caption{PageRank run-time (SK).}
\label{fig:back:process-vs-replication_sk-2005-2}
\end{subfigure}
\caption{Performance comparison of graph partitioning algorithms for PageRank computation on \emph{Friendster (FR)} and \emph{sk-2005 (SK)} graph.}
\label{fig:process-vs-replication}
\end{figure}

PageRank is executed on the graphs \emph{Friendster}~\cite{snap.datasets}  and \emph{sk-2005}~\cite{BCSU3, BoVWFI, BRSLLP} for 50 iterations on a Spark/GraphX cluster with 64 machines. \emph{Friendster} and \emph{sk-2005} consists of $1.8 \; \mathit{B}$ and $1.9 \; \mathit{B}$  edges and $66 \; \mathit{M}$ and $51 \; \mathit{M}$ vertices, respectively. The graphs are partitioned into 64 partitions with 2D~\cite{graphx}, 2PS~\cite{twophasestreaming} and NE~\cite{ne} as representatives for stateless streaming, stateful streaming and in-memory partitioning algorithms, respectively. Additionally, we use CRVC~\cite{graphx} as a baseline. In Figure~\ref{fig:process-vs-replication}, we report the replication factor, the partitioning run-time and the graph processing run-time. 
 
We observe that a better replication factor leads to a lower processing run-time. However, algorithms that yield a low replication factor impose a longer partitioning run-time. 
In-memory partitioning with NE is on both graphs much better than stateless streaming with 2D in terms of replication factor and graph processing run-time, but takes a much higher partitioning run-time. The graph partitioning run-time of stateful streaming with 2PS is between the run-time of 2D and NE, while the replication factor and the graph processing run-time for 2PS depends on the graph. On \emph{sk-2005}, 2PS is very close to NE and therefore also much better than 2D. On \emph{Friendster}, 2PS is much worse than NE and close to 2D. These examples show a correlation between replication factor and graph processing run-time. Therefore, the replication factor is an important partitioning feature for graph processing run-time prediction.

\subsection{Label Propagation}
\begin{figure}[t]
  \centering
  \begin{subfigure}[b]{0.32\linewidth}
      \centering
      \includegraphics[width=\linewidth]{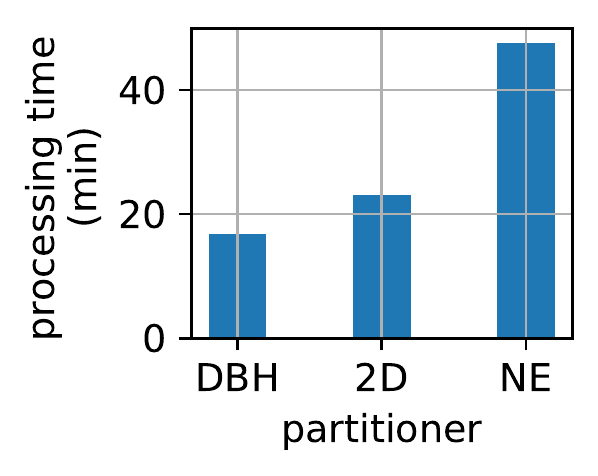}
      \vspace{-6mm}
      \caption{Label Propagation run-time (FB).}
      \label{fig:back:socfb-A-anon-processing-time}
  \end{subfigure}
  \hfill
  \begin{subfigure}[b]{0.32\linewidth}
      \centering
      \includegraphics[width=\linewidth]{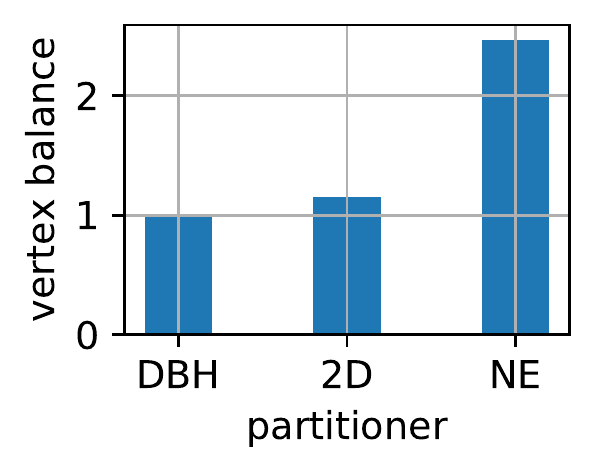}
      \vspace{-6mm}
      \caption{Vertex balance (FB).}
      \label{fig:back:socfb-A-anon-vertex-balance}
  \end{subfigure}
 \begin{subfigure}[b]{0.32\linewidth}
      \centering
      \includegraphics[width=\linewidth]{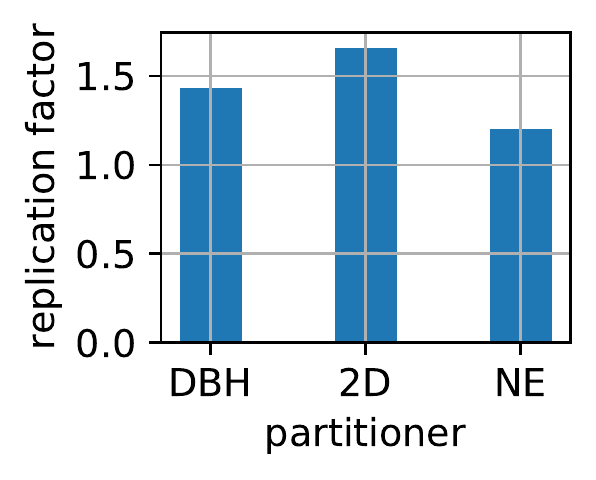}
      \vspace{-6mm}
      \caption{Replication factor (FB).}
      \label{fig:back:socfb-A-anon-replication-factor}
  \end{subfigure}
 
   \caption{Performance comparison of graph partitioning algorithms for Label Propagation computation on \emph{Socfb-A-anon (FB)} graph.}
   \label{fig:process-vs-balance}
   \vspace{-4mm}
\end{figure}
Label Propagation is executed on the graph \emph{Socfb-A-anon} \cite{network.datasets}  for 10 iterations on a Spark/GraphX cluster with 4 machines. \emph{Socfb-A-anon} consist of $3.1 \; \mathit{M}$ vertices and $24 \; \mathit{M}$ edges. The graphs are partitioned with 2D \cite{graphx}, DBH \cite{dbh} and NE \cite{ne} into 4 partitions. In Figure \ref{fig:process-vs-balance}, we report the graph processing run-time, the vertex balance and the replication factor. The graph partitioning run-time is not reported since it is negligible compared to the graph processing run-time in this scenario. 
We observe that a better vertex balance (i.e., close to $1.0$) leads to a lower processing run-time. The replication factor is less important since label propagation is computation-bound. We conclude that balancing metrics can be important features for graph processing run-time prediction.

These examples show that the prediction of graph processing run-time can depend on different partitioning metrics, and that partitioning run-time can also play a significant role in selecting a graph partitioner. We use these insights to design \GL, as described in the next section. 

\section{Approach}
\label{sec:approach}
\subsubsection*{Motivation}
\begin{figure}[t]
  \centering
  \includegraphics[width=\columnwidth]{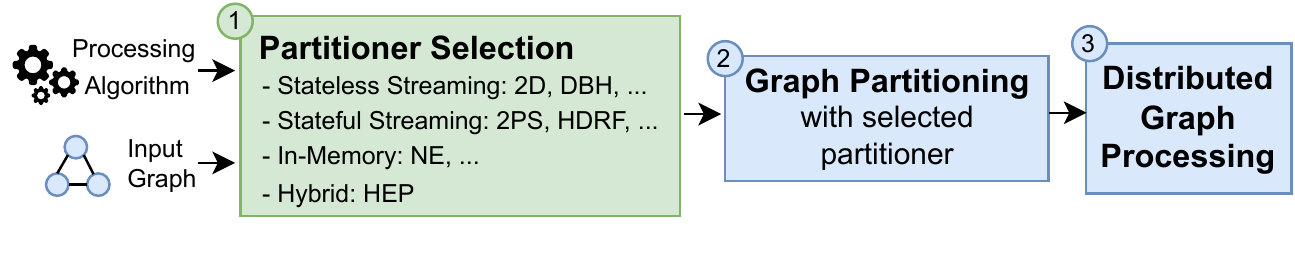}
  \vspace{-7mm}
  \caption{\GL {} is part of a graph processing pipeline and is used for automatic partitioner selection.}
  \label{fig:pipeline}
\end{figure}

Figure \ref{fig:pipeline} illustrates a graph processing pipeline consisting of three phases:
First, a graph partitioner is selected. Second, the graph is partitioned and finally, the distributed graph processing is performed. 
The partitioner selection is an important phase in this pipeline since it influences the run-time of the graph partitioning and graph processing phase. However, until now, the graph partitioner can only be selected with manual heuristics and best practices which make it a challenging task for unexperienced users. It is hard to tell which selection will lead to the lowest end-to-end run-time for a given graph processing workload and graph. Additionally, the selection phase needs to be really fast so that it does not slow down the pipeline. In order to tackle these challenges, we propose a machine learning-based system called \emph{\textbf{E}dge p\textbf{A}rtitioner \textbf{SE}lection}~(\GL{}) which enables efficient automatic partitioner selection.

\subsubsection*{Approach Overview}

\begin{figure}[t]
  \centering
  \includegraphics[width=\columnwidth]{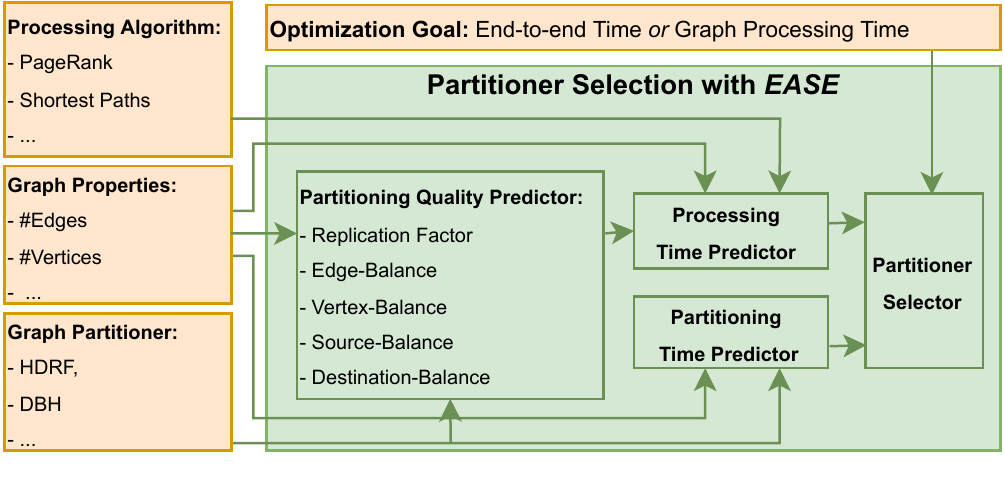}
  \vspace{-7mm}
  \caption{\GL {} - System Overview.}
  \label{fig:ease}
  \vspace{-6mm}
\end{figure}
\GLSS{} design is visualized in Figure~\ref{fig:ease}. 
The system consists of four components:
(1)~\QP, to  predict different partitioning quality metrics (replication factor, edge balance, vertex balance, destination balance and source balance) for different partitioners on a given graph. 
(2)~\PaP, to predict the partitioning run-time for different partitioners on a given graph.
(3)~\PrP, to predict the graph processing run-time of a given graph processing algorithm for a partitioned graph with the corresponding partitioning quality metrics.
(4)~\PS {} which, based on the prediction of partitioning and processing run-time, automatically selects a partitioner to either minimize the processing run-time or the end-to-end run-time. 

\subsubsection*{Training Phase Overview}
\begin{figure}[t]
  \centering
  \includegraphics[width=\columnwidth]{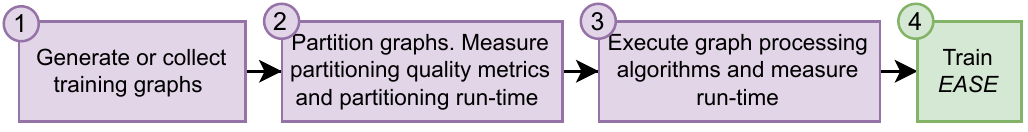}
  \vspace{-5mm}
  \caption{The training phase of \GL {} consists of four steps.}
  \label{fig:training}
  \vspace{-6mm}
\end{figure}
The training phase of \GL {} consists of four steps (see Figure \ref{fig:training}). First, graphs with different properties are collected or generated. Second, the graphs are partitioned with different partitioning algorithms into different numbers of partitions and the partitioning quality metrics and the partitioning run-time are measured. Third, different graph processing algorithms are executed for each combination of graph and partitioner and the processing run-time is measured. Forth, the acquired data is used to train machine learning models for \QP, \PaP {} and \PrP. Finally, the models can be applied to enable automatic partitioner selection. 

In the following, we show how to acquire enough training data which covers a wide spectrum of graphs with different properties (Section~\ref{sec:training_data_aquisition}) and which graph properties to select as features for the machine learning models (Section~\ref{sec:machine_learning_features}). Afterwards, we describe how to select and fine tune the machine learning models (Section~\ref{sec:training}) and how to apply the models to a given graph problem (Section~\ref{sec:inference}). Finally, we discuss design decisions for \GL {} (Section \ref{sec:designchoices}).

\subsection{Training Data Acquisition}
\label{sec:training_data_aquisition}
The goal of data acquisition is to cover a wide spectrum of graph properties of unseen real-world graphs that are expected as a workload and for which the three prediction tasks should be performed. At the same time, it should be easy to obtain these graphs. In Section \ref{sec:indroduction}, we described that it can be challenging for several reasons to acquire enough real-world graphs. We tackle this challenge by using synthetic graphs. By doing so, we can generate many graphs and cover a wide spectrum of graph properties. This is our basis to build machine learning models. Then, the models are tested on real-world graphs and possible weaknesses can be identified, e.g., that certain combinations of graph types and partitioning algorithms do not yield accurate results. For these weaknesses of the model, the training can be refined by enriching the synthetic data with real-world graphs, if available.

In our work, we use R-MAT \cite{rmat} as graph generator and use the implementation of Khorasani et al. \cite{rmat.generator}. R-MAT is lightweight, scales well, produces realistic graphs \cite{Chakrabarti.survery, bonifati.generator, TrillionG} and is used in the well-known Graph500 benchmark \cite{introducinggraph500}. According to the original R-MAT paper \cite{rmat}, the partitions $a$ and $d$ can be seen as communities that are connected by edges in the partitions $b$ and $c$. The recursive generation process leads to sub-communities. In order to generate different graphs, we choose different combinations of parameters. First, we use different number of vertices ($2^{15}$ to $2^{27}$) and edges (1 M to 200 M). All combinations are shown in Table \ref{tab:rmat-small-vertices}. Thereby, we can systematically generate graphs with different average degrees and densities (which are seen in graph repositories like SNAP \cite{snap.datasets} and KONECT \cite{connect.datasets}). Second, for each combination of numbers of vertices and edges, we use nine combinations of values for $a$, $b$, $c$ and $d$. By doing so, we influence the skewness of the degree distribution, the average local clustering coefficient and the mean number of triangles. For all graphs, we fixed $d$ to 5\%. We use 19\% or 34\% for $c$, which leads to different numbers of edges between community $a$ and $d$. Now, we use different combinations for $a$ and $b$. Larger values for $a$ lead to more skew in the degree distribution and a larger community $a$. Larger values for $b$ lead to more edges between $a$ and $d$ (inter-cluster edges). All nine R-MAT parameter combinations are shown in Table \ref{tab:rmatparams}. 
  
\begin{figure*}[t]
     \centering
     \begin{subfigure}[b]{0.15\linewidth}
         \centering
         \includegraphics[width=\linewidth]{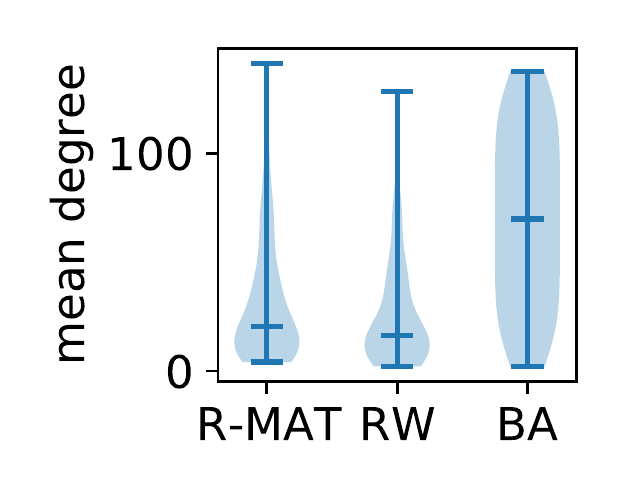}
         \vspace{-8mm}
         \caption{}
         \label{fig:feature-distributions:deg}
     \end{subfigure}
     \hfill
     \begin{subfigure}[b]{0.15\linewidth}
         \centering
         \includegraphics[width=\linewidth]{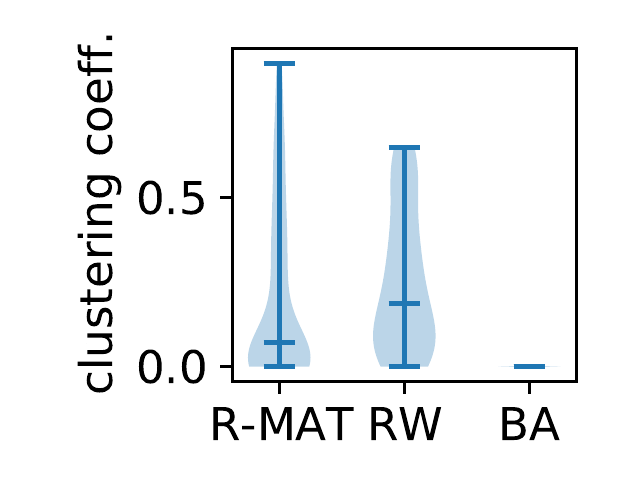}
         \vspace{-8mm}
         \caption{}
         \label{fig:feature-distributions:lcc}
     \end{subfigure}
     \hfill
     \begin{subfigure}[b]{0.15\linewidth}
         \centering
         \includegraphics[width=\linewidth]{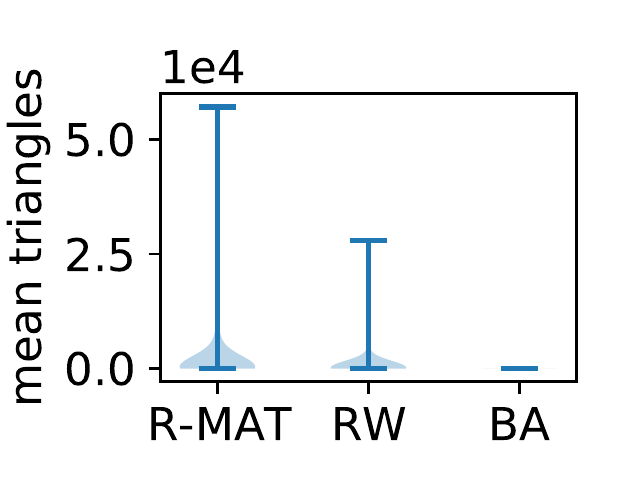}
         \vspace{-8mm}
         \caption{}
         \label{fig:feature-distributions:triangles}
     \end{subfigure}
   \hfill
     \begin{subfigure}[b]{0.15\linewidth}
         \centering
         \includegraphics[width=\linewidth]{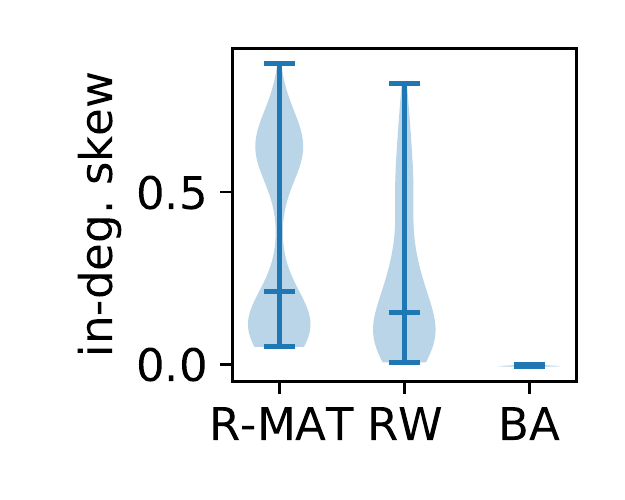}
         \vspace{-8mm}
         \caption{}
         \label{fig:feature-distributions:degin}
     \end{subfigure}
   \hfill
     \begin{subfigure}[b]{0.15\linewidth}
         \centering
         \includegraphics[width=\linewidth]{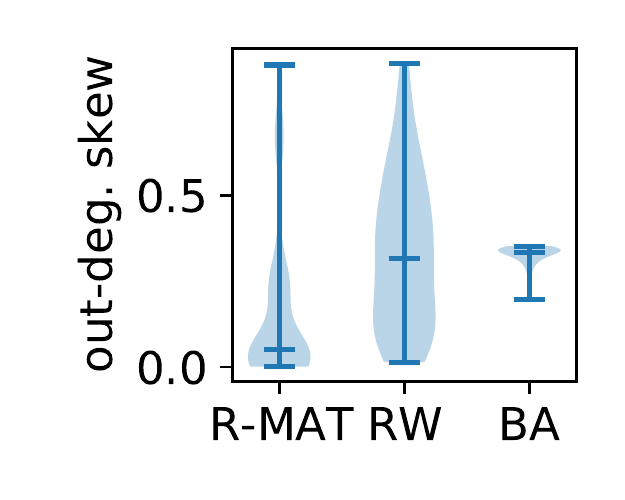}
         \vspace{-8mm}
         \caption{}
         \label{fig:feature-distributions:degout}
     \end{subfigure}
      \hfill
     \begin{subfigure}[b]{0.22\linewidth}
         \centering
         \includegraphics[width=\linewidth]{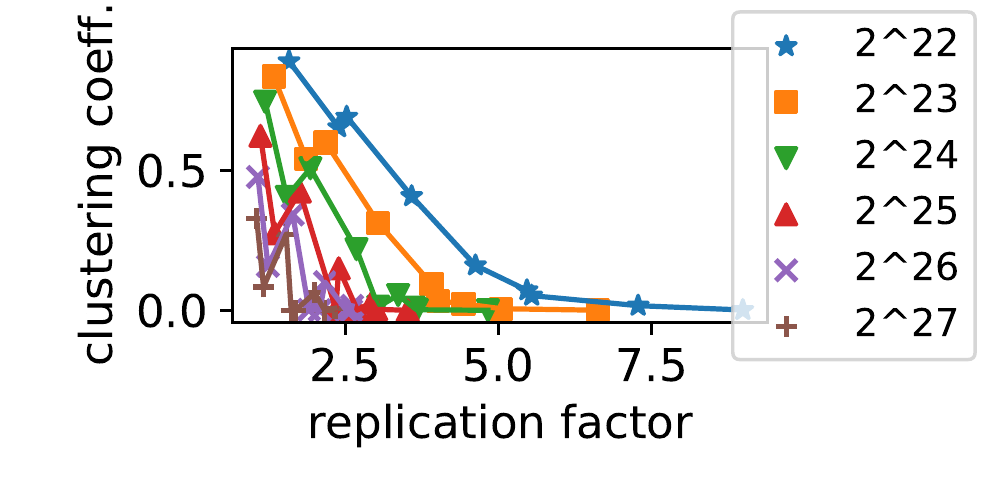}
         \vspace{-8mm}
        \caption{}
         \label{fig:feature-distributions:rep-vs-cc}
     \end{subfigure}

     \hfill
     \vspace{-5mm}
      \caption{(a)-(e): Graph properties of generated R-MAT and Barabasi-Albert (BA) graphs and real-world (RW) graphs, (f)~correlation between high clustering coefficients and low replication factors.}
      \label{fig:feature-distributions}
      \vspace{-5mm}
  \end{figure*}

We compare the graph properties of the 297 generated graphs with real-world graphs. The violin plots in Figures~\ref{fig:feature-distributions:deg} to \ref{fig:feature-distributions:degout} show the distribution of the graph property values along with the minimal, maximal and median value of the real-world and generated graphs. The graph properties of real-world graphs are covered by the generated graphs to a large extent, which indicates good coverage of real-world graph properties. 

In Figure \ref{fig:feature-distributions:rep-vs-cc}, it is shown that the different combinations of values for $a$, $b$, $c$ and $d$ influence the clustering coefficient of the generated graphs and how ``easy'' the graphs can be partitioned. In the figure, multiple graphs are represented as follows. All graphs have $|E|=160 \, \text{M}$ edges. Each line in the diagram represents how many vertices $n \in \{ 2^{22}, 2^{23}, 2^{24}, 2^{25}, 2^{26}, 2^{27}\}$ the graph contains. The markers on the lines are the nine different parameter combinations of values for $a$, $b$, $c$, and $d$ and therefore represent one single graph each.  On the y-axis, the clustering coefficient of the graphs is shown. On the x-axis, the replication factor achieved by partitioning the graph with the streaming partitioner \emph{High-Degree Replicated First (HDRF)} \cite{hdrf} into $64$ partitions is shown. We observe a correlation between high clustering coefficients and low replication factors. Similar results were observed for other graphs and partitioning algorithms. 
  
The 297 R-MAT graphs are used to train \QP.  However, the graphs are too small to measure representative figures of the performance of distributed graph processing. Hence, we additionally generated 180 larger graphs with the same parameter combinations for the R-MAT generator from above (see Table~\ref{tab:rmatparams}). These graphs contain $1.8 \, \text{M}$ to $50 \, \text{M}$ vertices and $100 \, \text{M}$ to $500 \, \text{M}$ edges (see Table~\ref{tab:rmat-large-vertices}). The larger graphs are used to train both \PrP {} and \PaP {}.

We also attempted to use the Barabasi-Albert model \cite{barabasi} as graph generator (implementation of Hagberg et al. \cite{networkx}). The generator expects the total number of vertices $|V|$ and the number of edges $m$ which are added for each new vertex. We created 70 graphs with $|V|=1 \, \text{M} $ and $m \in \{1,2, \dots, 70\}$ to obtain graphs with average degrees between $2$ and $140$. The average degree influences the replication factor). 
However, if we fix $m$ and change $|V|$, the replication factor does not change for all partitioning algorithms. Therefore, it is not possible to create graphs with the same average degree that lead to different replication factors. 
In contrast to R-MAT, it is also not possible to cover all the graph properties of real-world graphs.

In conclusion, using R-MAT for graph generation yields promising results, while the Barabasi-Albert model is not flexible enough to generate a broad variety of graphs. 
  
\begin{table}[t]
  \caption{Different combinations of the number of edges and vertices in the generated R-MAT graphs used for training of (a) \QP {} and (b) \PrP {} and \PaP.}\label{tab:rmat-vertices}
  \centering
  \begin{subtable}[t]{0.58\columnwidth}
    \caption{R-MAT-SMALL}\label{tab:rmat-small-vertices}
    \centering
    \begin{tabular}{|>{\scriptsize}l|>{\scriptsize}l|}
      \hline
      $|E|$ (M) & $|V|$  \\
      \hline\hline
    1  & $2^{15}$, $2^{16}$, $2^{17}$, $2^{18}$, $2^{19}$ \\
    40  & $2^{21}$, $2^{22}$, $2^{23}$, $2^{24}$, $2^{25}$\\
    80  & $2^{21}$, $2^{22}$, $2^{23}$, $2^{24}$, $2^{25}$, $2^{26}$ \\
    120  & $2^{22}$, $2^{23}$, $2^{24}$, $2^{25}$, $2^{26}$ \\
    160  & $2^{22}$, $2^{23}$, $2^{24}$, $2^{25}$, $2^{26}$, $2^{27}$ \\
    200  & $2^{22}$, $2^{23}$, $2^{24}$, $2^{25}$, $2^{26}$, $2^{27}$ \\
       \hline
    \end{tabular}
  \end{subtable}
  \hfill
  \begin{subtable}[t]{0.40\columnwidth}
    \caption{R-MAT-LARGE}\label{tab:rmat-large-vertices}
    \centering
    \begin{tabular}{|>{\scriptsize}l|>{\scriptsize}p{2cm}|}
      \hline
      $|E|$ (M) & $|V|$ (M)  \\
      \hline\hline
    100 & 1.8, 2.5, 4, 10 \\
    200 & 3.6, 5, 8, 20 \\
    300 & 5.4, 7.5, 12, 30 \\
    400 & 7.3, 10, 16, 40\\
    500 & 9.1, 12.5, 20, 50\\
       \hline
    \end{tabular}
  \end{subtable}
\end{table}

  \begin{table}[t]
    \caption{Nine combinations $\mathbf{C_i}$ for the R-MAT parameters a, b, c and d used for generating training graphs.}
    \vspace{-2mm}
    \label{tab:rmatparams}
    \begin{center}
      
    \begin{tabular}{|>{\scriptsize}c|>{\scriptsize}c|>{\scriptsize}c|>{\scriptsize}c|>{\scriptsize}c|>{\scriptsize}c|>{\scriptsize}c|>{\scriptsize}c|>{\scriptsize}c|>{\scriptsize}c|}
    \cline{2-10} 

    \multicolumn{1}{c|}{} & $\mathbf{C_1}$ & $\mathbf{C_2}$  & $\mathbf{C_3}$  &  $\mathbf{C_4}$  &  $\mathbf{C_5}$  & $\mathbf{C_6}$   &  $\mathbf{C_7}$   &  $\mathbf{C_8}$  &  $\mathbf{C_9}$\\
 
    \hline
    \textbf{a} & 0.35 &0.45 & 0.55 & 0.60&    0.40 &0.50 & 0.60 & 0.65 & 0.70\\
    \hline
    \textbf{b} & 0.26 &0.16 & 0.06 & 0.01&    0.36 &0.26 & 0.16 &0.11 & 0.06\\
    \hline
    \textbf{c} &  0.34 &0.34 &0.34 & 0.34&    0.19 &0.19 & 0.19 & 0.19 & 0.19\\
      \hline
      \textbf{d} & 0.05 &0.05  & 0.05  & 0.05 &    0.05  &0.05  & 0.05 & 0.05 & 0.05 \\
     \hline
  
  \end{tabular}
  \end{center}
  \vspace{-6mm}
  \end{table}

\subsection{Machine Learning Features}
\label{sec:machine_learning_features}
For all three prediction components of \GL{}, we use two kinds of features. (1)~Common \emph{graph properties} in order to map the graph to the closest graph(s) we have already processed and (2)~\emph{other features} which are specific to the prediction component.

\subsubsection{Graph Properties}
In total we use three different sets of graph properties as features (see Table~\ref{tab:graph_features}). In the following, we show which features are used for which prediction task.

\QP {} uses the basic and advanced graph properties as features. The rationale for using two different feature sets is that the basic graph properties can be easily computed if not known, compared to the advanced graph properties which are more compute-intensive to obtain. However, the advanced features provide more insights into the characteristics of the graph and therefore may lead to a better prediction. In many graph dataset repositories~\cite{snap.datasets, connect.datasets, network.datasets}, advanced properties are precomputed.
The basic features are the \emph{mean degree} of the graph, the \emph{skewness of the in-degree distribution}, the \emph{skewness of the out-degree distribution} and the \emph{density} of the graph. These are common properties of graphs, and it is expected that they provide insights about how well a graph can be partitioned. In experimental studies~\cite{survey.1.vldb.2017, survey.4.sigmod.2019}, the degree distribution was considered for the selection of the partitioner. Furthermore, it was observed that different partitioners can be sensitive to the skewness of the degree distribution \cite{hdrf}.  
The advanced features are, in addition to the basic features, the \emph{average number of triangles} and the \emph{average local clustering coefficient}. It was observed that many real-world networks have a community structure which leads to a high clustering coefficient~\cite{watts}. This can be an interesting indicator for how well a graph can be partitioned. In~\cite{zeng}, the authors observed that their partitioner performed better (in terms of cut-size) on graphs with higher clustering coefficients. Therefore, we use the advanced features to investigate how much they can improve the replication factor prediction. However, we have not found any literature that shows the advanced features would influence the balancing. Therefore, we use the advanced features only for the replication factor prediction and not for balancing prediction.  

\PrP {} only considers the simple features. Many graph processing algorithms perform computations for each edge or vertex. Therefore, the number of edges and vertices are the most important graph properties.

For the \PaP {}, all feature sets are considered. For the stateless partitioners, the simple features will be most important as they represent the size of the graph. However, for the more advanced partitioners, also the basic and advanced properties can be important features. For example, in HEP~\cite{hep}, the decision of how much of the graph is partitioned in-memory and how much in a streaming fashion is decided based on the mean degree. Therefore, the mean degree influences the partitioning-runtime~\cite{hep}. As another example, 2PS performs clustering as a pre-processing step and sorts the clusters~\cite{twophasestreaming}. Therefore, the clustering coefficient can be an important feature.

\subsubsection{Other features}
\label{approach:other-features}
In addition to the graph properties, we consider features which are specific to the prediction component. 

\QP {} always considers as specific features the graph partitioner and the number of partitions. Some graph partitioners use a configuration parameter that influences the partitioning quality. We treat different settings of partitioner-specific parameters as if they were separate partitioners. We showcase this in Section~\ref{sec:evaluation} for the HEP partitioner~\cite{hep} by using different parameters for $\tau \in \{1,10,100\}$ that influence how many edges of the graph are partitioned in-memory and how many in a streaming fashion. The three settings result in three different partitioners HEP-1, HEP-10 and HEP-100. 

\PrP {} uses the partitioning quality metrics (replication factor, edge balance, vertex balance, source balance and destination balance) as specific features. For each graph processing algorithm, a separate model is trained. Thereby, new graph processing algorithms can easily be added and trained individually.   

\PaP {} considers the partitioner as specific feature. As for the quality prediction, we treat different settings of partitioner-specific parameters as if they were separate partitioners. 

\begin{table*}[t]
  \caption{Overview which graph properties are used in which feature set and which features are used for which prediction component. I is the number of iterations of a given graph processing algorithm (e.g., PageRank), k is the number of partitions.}
  \vspace{-2mm}
  \label{tab:graph_features}
  \begin{center}
\begin{tabular}{|>{\scriptsize}c|>{\scriptsize}c|>{\scriptsize}c|>{\scriptsize}c|>{\scriptsize}c|>{\scriptsize}c|>{\scriptsize}c|>{\scriptsize}c|>{\scriptsize}c|>{\scriptsize}c|>{\scriptsize}c|>{\scriptsize}c|>{\scriptsize}c|>{\scriptsize}c|}
  \cline{2-14}
  \multicolumn{1}{c}{} &\multicolumn{8}{|c|}{\textbf{Graph Properties (Sec. \ref{sec:background:graphproperties})} } &\multicolumn{5}{|c|}{ \textbf{Other Features (Sec. \ref{approach:other-features})}}\\
  \cline{1-14}
   \textbf{Description} & $\mathbf{|E|}$ & $\mathbf{|V|}$ &\shortstack{\textbf{Mean} \\ \textbf{Degree}}&  \textbf{Density} &  \shortstack{\textbf{In-Deg.} \\ \textbf{Dist.}}  & \shortstack{\textbf{Out-Deg.} \\ \textbf{Dist.}}  & \shortstack{\textbf{Average} \\ \textbf{\#triangles}}  & \textbf{LCC} & \shortstack{\textbf{Quality}  \textbf{Metrics} \\ \textbf{(Sec. \ref{sec:background:partitioningqualitymetrics})}}   & \textbf{k} & \shortstack{\textbf{Partitioner}}  & \textbf{I} & \shortstack{\textbf{Opt.} \\ \textbf{Goal}} \\ 
    \hline
    \multicolumn{14}{c}{Feature Sets}\\   
    \hline
    \textbf{Simple} &  \cmark &  \cmark &   &  &  &  &  &   & && &  & \\ 
    \hline
    \textbf{Basic} &  \cmark &  \cmark &  \cmark & \cmark  &\cmark & \cmark &   &  & & & & & \\
    \hline
    \textbf{Advanced} &  \cmark &  \cmark &  \cmark & \cmark  &\cmark & \cmark & \cmark  & \cmark & & &&  & \\
    \hline
    \multicolumn{14}{c}{Prediction Tasks and Partitioner Selector}\\   
    \hline
    \shortstack{\textbf{Partitioning Quality}} & &   &  \cmark & \cmark  & \cmark  & \cmark  & \cmark  & \cmark & & \cmark &\cmark &  & \\
    \hline
    \shortstack{\textbf{Partitioning Time}} &  \cmark &  \cmark &  \cmark & \cmark  & \cmark  & \cmark  &  &  & & &\cmark &  & \\
    \hline
    \shortstack{\textbf{Processing Time}} &  \cmark &  \cmark &   &   &  &   &   &  & \cmark & &  & \cmark & \\
  \hline\hline
    \shortstack{\textbf{Selector}} &  \cmark &  \cmark &  \cmark & \cmark  & \cmark  & \cmark  & \cmark  & \cmark & \cmark &\cmark & \cmark & \cmark &\cmark \\
    \hline
\end{tabular}
\end{center}
\vspace{-5mm}
\end{table*}

\newpage
\subsection{Training}
\label{sec:training} 
For each of the three prediction components, we compare six supervised machine learning algorithms.
\emph{Polynomial Regression} is effective in predicting computational costs in practice~\cite{application.driven.partitioning}. 
\emph{Support Vector Regression~(SVR)} leads to a very good performance and is considered as an important technique for regression tasks~\cite{svrarg,svrargg}. 
\emph{Random Forest Regressor~(RFR)} is relatively robust to outliers and noise, can easily be parallelized and the importance of the used features can be interpreted~\cite{breiman2001random} - we leverage the interpretability of RFR in Section~\ref{sec:feature_importance}.
\emph{Extreme Gradient Boosting~(XGB)} leads to state-of-the-art results and is commonly used in well-known machine learning competitions~\cite{xgbscale}. Decision tree ensembles such as RFR or XGB can compete and even outperform deep neural networks on \emph{tabular data}~\cite{9998482, arxiv220708815, gorishniy2021revisiting, shwartz2022tabular}. \emph{K-nearest Neighbors Regressor~(KNN)} serves as a simple baseline. 
In addition to these traditional machine learning methods, we use a \emph{fully-connected multi-layer perceptron~(MLP)} as a representative for a feed-forward deep neural network.

\subsubsection*{Data Splitting}
\label{sec:approach:splitting}
All models are trained on the synthetically generated R-MAT graphs. \QP {} is trained on the smaller 297 R-MAT graphs and \PaP {} and \PrP {} are trained on the larger 180 R-MAT graphs. The hyper-\allowbreak parameter tuning is performed for each machine learning algorithm on the training set by using 5-fold cross-\allowbreak validation \cite{mohri2012foundations}. The models are compared against each other based on the cross-\allowbreak validation score and are retrained with the best hyper-\allowbreak pa\allowbreak ra\allowbreak meter on the complete training set. Finally, the models are evaluated on the test set, which consists only of real-world graphs.
\subsubsection*{Preprocessing}
We preprocess the generated data as follows: The data was standardized with z-score normalization~\cite{kreyszig2010advanced}. For \QP {} and \PaP, one-hot encoding is used for the partitioning algorithms to use them as a numerical feature. 

\subsubsection*{Hyper-Parameter Search}
\label{sec:approach:hyperparameter}
For all six machine learning algorithms, hyper-parameters can be tuned. 
We performed a grid search for each machine learning algorithm to identify the best hyper-parameters. 
Please find the used combinations in our GitHub repository \cite{graphlearner}.

\subsubsection*{Test \& Enrichment}
As the last step, our models are tested on real-world graphs. In order to identify weaknesses of \QP {}, the prediction performance is evaluated for different combinations of partitioning algorithm and graph type. If one combination does not perform well, but is relevant for the user, the training set can be enriched with \emph{real-world} graphs. For example, if a user mainly processes wiki graphs, but the predictions for this graph type are not accurate, the user can enrich the training data with real-world wiki graphs in order to improve the prediction performance.

\subsection{Inference}
\label{sec:inference}
In order to make predictions for an unseen graph, for \QP {}, the graph properties, the partitioner and the number of partitions need to be provided. For \PrP, the graph processing algorithm to execute, the graph properties and the partitioning quality metrics need to be provided. In contrast to graph processing algorithms that run until convergence, for graph processing algorithms that run for a given number of iterations, also the number of iterations needs to be provided. For \PaP, the graph properties and the partitioner is needed for inference. In order to automatically select a partitioner with \PS, the optimization goal to either minimize the processing or the end-to-end time, needs to be provided. Table~\ref{tab:graph_features} shows which features are needed for which prediction task.

\subsection{Discussion of Design Choices}
\label{sec:designchoices}
\textit{Seperate Machine Learning Models for Prediction Tasks:}
We discuss two alternatives to our approach. First, instead of predicting the partitioning and processing run-time separately, it is possible to train a single model which directly predicts the end-to-end time (Alternative~1). Second, for the processing run-time prediction, the partitioner could be used as a feature (Alternative~2) \emph{instead} of the partitioning quality metrics. Therefore, in Alternative~2 it is not necessary to predict partitioning quality metrics. 

We decided against Alternative~1 for two reasons. First, it is less flexible than our approach because only the end-to-end run-time can be predicted. Some users may only be interested in minimizing the graph processing run-time, e.g., when partitioning can be performed offline on a cheap compute node, while processing is running on an expensive compute cluster, and therefore, only processing costs should be minimized. Second, if a new partitioner is added or the processing framework is changed, the entire model needs to be retrained. With our approach, only the model for the partitioning run-time \emph{or} the processing run-time prediction needs to be retrained in such a scenario. 

We decided against Alternative~2 for the following reasons. First, with our approach, new partitioning algorithms can be incorporated with less overhead: If a new partitioner should be included, it simply needs to be executed on the training graphs to measure the quality metrics and the partitioning run-time. Then, \QP {} and \PaP {} can be retrained. \PrP {} does  \emph{not} need to be retrained, since a model already exists that predicts the graph processing run-time based on the quality metrics. In contrast, with Alternative~2, all graph processing algorithms would need to be re-executed on the training graphs, since the partitioner and not the partitioning quality is used as a feature. Second, with our approach, additional insights can be gained on how the partitioning quality metrics influence the processing run-time. This can be useful for researchers to better understand for which graph processing algorithm which quality metrics are most important. This insight may be useful for the design of new partitioning algorithms or optimizations of graph processing frameworks. Third, the replication factor prediction is also useful for memory-bound processing as each replica produces a copy of the vertex state. For example, in graph neural networks (GNNs), the vertex state can consist of thousands of features and memory is heavily restricted because the processing is performed on GPUs. In such a scenario, the predicted replication factor can be an important decision criterion for selecting a partitioner.

\textit{Use of Graph Embeddings as Features:}
It would also be possible to use complex graph features for the prediction tasks by computing graph embeddings with graph neural networks~(GNNs) to represent a graph with low dimensional vectors \cite{8294302, NIPS2017_5dd9db5e, kipf2017semi, 4700287}. In our experiments\footnote{In our GitHub repository, we report all inference times on a CPU and two different GPUs and compare them with EASE. Here, we summarize these results.}, we measure the inference time of a two-layer GraphSage GNN \cite{NIPS2017_5dd9db5e} with mean pooling on a CPU and on two different GPUs for all graphs of Table~\ref{tab:real-world-graphs-processing}. We observe that the inference time on the CPU is 9 to 19 times the partitioning time of the \emph{slowest} partitioner. Therefore, it is possible to run all partitioners and to measure the actual metrics instead of predicting them, so we could not amortize the high inference time end-to-end. Inference on GPUs at this scale is heavily bounded by GPU memory. Even on a large GPU with 48~GB GPU memory, we can only compute embeddings for the first four graphs of Table~\ref{tab:real-world-graphs-processing} and run out of GPU memory for the remaining graphs. It will not pay off to use such heavy-weight processing just to optimize the partitioning step of distributed graph processing. Therefore, we do not use GNNs in our system.

\section{Evaluation}
\label{sec:evaluation}
In the following, we evaluate the ability of \GL{} to predict the partitioning quality metrics, the partitioning run-time and the processing run-time.
We also compare \GL{} to manual heuristics in selecting a graph partitioner, such as random selection or selection of the partitioner that yields to the lowest replication factor. 
\subsection{Evaluation Metrics}
We evaluate the machine learning models using evaluation metrics that are commonly used for regression tasks. In the following formulas, $y_i$ is the true value (e.g., the replication factor) for the i-th sample, $\hat{y_i}$ is the predicted value for the i-th sample, $\overline{y}$ is the mean value of all samples and $\epsilon$ is a small value in order to prevent divisions by zero.
\begin{enumerate} 
  \item Root Mean Squared Error is defined as \\ $\mathit{RMSE}(y, \hat{y}) = \sqrt[2]{\frac{1}{n} \sum_{i=0}^{n-1} (y_i - \hat{y_i} )^2}$. The closer the value is to 0, the better.
  \item Mean Absolute Percentage Error is defined as \\ $\mathit{MAPE}(y, \hat{y}) \allowbreak = \frac{1}{n} \sum_{i=0}^{n-1} \frac{|y_i - \hat{y_i}|}{max(\epsilon, |y_i|)}$. The closer the value is to 0, the better. 
\end{enumerate}
\subsection{Test Set}
\label{sec:testset}
\begin{table}[t]
  \caption{Test set (real-world graphs) for \PaP {} and \PrP.}
  \vspace{-2mm}
  \label{tab:real-world-graphs-processing}
  \begin{center}
    \begin{tabular}{|>{\scriptsize}l|>{\scriptsize}l|>{\scriptsize}l|}
      \hline
                            \textbf{Graph} &  \textbf{Edges (M)} &  \textbf{Vertices (M)} \\
                            \hline\hline
                com-orkut.ungraph &      117.2 &           3.1 \\
                      enwiki-2021 &      150.1 &           6.3 \\
                      eu-2015-tpd &      165.7 &           6.7 \\
                   hollywood-2011 &      229.0 &           2.0 \\
       out.orkut-groupmemberships &      327.0 &           8.7 \\
                     eu-2015-host &      379.7 &          11.3 \\
                     gsh-2015-tpd &      581.2 &          30.8 \\
                     \hline
      \end{tabular}
\end{center}
\vspace{-6mm}
\end{table}
For the evaluation of \QP, we collected 175 real-world graphs from SNAP~\cite{snap.datasets}, KONECT~\cite{connect.datasets} and the Network Data Repository~\cite{network.datasets}. The graphs can be categorized into the following graph types: (1)~12 affiliation graphs, (2)~3 citation graphs, (3)~6 collaboration graphs, (4)~5 interaction graphs, (5)~5 internet graphs, (6)~one product network, (7)~31 social networks, (8)~12 web graphs, (9)~101 wiki graphs. All graphs except 96 of the 101 wiki graphs are used as the final test set for all evaluations. The 96 wiki graphs are later used to evaluate training data enrichment. From the same graph data repositories, we collected 7 real-world graphs for the evaluation of \PaP {} and \PrP {} (see Table~\ref{tab:real-world-graphs-processing}). The links to all graph datasets are provided in our GitHub repository~\cite{graphlearner}.
\subsection{Training, Validation \& Test}
\subsubsection*{Data Generation}
For all three prediction components we measure the graph properties introduced in Section~\ref{sec:background:graphproperties} of the training and the test graphs. 
We use 11 different partitioners of four categories. (1)~Stateless streaming with 1-dimensional hashing of destination vertices~(1DD) and source vertices~(1DS)~\cite{graphx}, 2-dimensional hashing~(2D)~\cite{graphx}, canonical random vertex cut~(CRVC)~\cite{graphx} and degree-based hashing~(DBH)~\cite{dbh}. (2)~Stateful streaming with high-degree replicated first~(HDRF)~\cite{hdrf} and two phase streaming~(2PS)~\cite{twophasestreaming}. (3)~In-memory partitioning with Neighborhood Expansion~(NE)~\cite{ne}. (4)~Hybrid partitioning with Hybrid Edge Partitioner~(HEP)~\cite{hep}. HEP uses a parameter $\tau$ to control how many edges are partitioned in-memory and how many in a streaming fashion~\cite{hep}. As the authors suggest, we configured HEP with $\tau \in \{1, 10, 100\}$ abbreviated as HEP-1, HEP-10, HEP-100 and treat them as different partitioners. Therefore, we have 11 partitioners in total. The selected partitioning algorithms are strong baselines in their respective category and claim to have a good balance in terms of partitioning quality and partitioning run-time. Therefore, all of them are potential candidates to minimize end-to-end run-time and it is difficult to manually decide on the optimal partitioning strategy.   

For \QP {}, we partition the graphs into $k \in K = \{4, 8, 16, 32, 64, 128\}$ partitions and measure the five partitioning quality metrics discussed in Section \ref{sec:background:partitioningqualitymetrics}. 

For \PrP, we partition all graphs with all partitioners into 4 partitions and measure the partitioning quality metrics. Then, we execute 6 graph processing algorithms on a Spark/GraphX cluster with 4 machines: PageRank~(PR) for 10 iterations, Connected Components~(CC), Single Source Shortest Paths~(SSSP) with a randomly selected seed vertex and K-Cores with $k=deg(G)$. These algorithms are characteristic for different sorts of workloads. In PageRank, all vertices are active in each iteration. In Connected Components, in the first iteration all vertices are active and the number of active vertices decreases over time. In Single Source Shortest Paths, in the first iteration only one vertex is active. In the following iterations, the number of active vertices first increases and then decreases until no vertex is active anymore. In K-cores, in the first iteration many vertices are active and become inactive over time. In addition to these 4 real-word algorithms, we implement a synthetic algorithm: Each vertex in the graph contains a feature vector with $s$ 64-bit doubles and sends its feature vector along the outgoing edges in each iteration. With $s$, the amount of communication can be influenced. We set $s$ to 1 and to 10 and abbreviate these two configurations with Synthetic-Low and Synthetic-High for low and high communication load, respectively. We set the number of iterations to 5. 
For both the synthetic workloads and PageRank, the computation and communication load remains constant throughout the iterations and the number of iterations is a parameter of the algorithms. Therefore, we measure the average iteration time which is also the prediction target. The total processing run-time is the average iteration time multiplied with the number of iterations. For the remaining algorithms in which the computation and communication load changes over time and which run until convergence, we measure the total processing time until convergence. 

For the \PaP, we partition all graphs with all partitioners into 4 partitions and measure the partitioning run-time. 

\subsubsection*{Training and Validation}
For all three prediction components, we train regression models with all six machine learning algorithms mentioned in Section~\ref{sec:training} and select the best model based on the 5-fold cross-validation score on the synthetic graphs. 
In the following, we evaluate the selected models on the test set consisting of real-world graphs.
\subsubsection*{Test}
We evaluated \PaP {} and \PrP {} on the seven real-world graphs listed in Table \ref{tab:real-world-graphs-processing}. 
For \PaP {} we achieved a MAPE of 0.335 on the test set with XGB. For \PrP, XGB led to a MAPE of 0.295, 0.401, 0.300 for PageRank, K-Cores and Single Source Shortest Paths, respectively (see Table~\ref{tab:scores-processing}). For Connected Components, Synthetic-Low and Synthetic-High, Polynomial Regression led to a MAPE of 0.272, 0.271, 0.259, respectively.
\begin{table}[t]
  
  \caption{Quality metrics for \PrP {} on test set (average across real-world graphs).} 
  \vspace{-2mm}
  \label{tab:scores-processing}
  \begin{center}
  \begin{tabular}{|>{\scriptsize}l|>{\scriptsize}l|>{\scriptsize}l|>{\scriptsize}l|>{\scriptsize}l|>{\scriptsize}l|>{\scriptsize}l|}
     \hline
     \textbf{Algorithm} & \textbf{Model} & \textbf{MAPE}    \\
    \hline\hline
    Connected Components & PolyRegression & 0.272 \\
    K-Cores & XGB & 0.401\\
    PageRank & XGB &  0.295\\
    Single Source Shortest Paths & XGB & 0.300 \\
    Synthetic-High & PolyRegression & 0.259 \\
    Synthetic-Low & PolyRegression & 0.271 \\
     \hline
\end{tabular}
\end{center}
\vspace{-6mm}
\end{table}
We evaluated \QP {} with the 80 real-world graphs mentioned in Section~\ref{sec:testset}. Table~\ref{tab:scores-std} shows the results. RFR led to a MAPE of 0.152, 0.144, 0.079, 0.154 for vertex balance, source balance, edge balance and destination balance, respectively. In terms of replication factor, XGB led to a MAPE of 0.296 for the basic features and could be slightly improved to 0.288 by using the advanced features. This means that the balancing metrics can be predicted more accurately than the replication factor. We further investigate for which combination of graph type and partitioning algorithm the model works well and for which combinations further training would be useful. The results for the replication factor are shown in Figure~\ref{fig:eval:rfr-basic-0-heatmap}. We observe that all partitioning algorithms achieve good scores for the graph types \textit{citation}, \textit{interaction}, \textit{internet}, \textit{product-network} and \textit{social}. However, the predictions for the partitioners 2PS, HDRF, HEP and NE on the graph types \textit{collaboration}, \textit{web} and \textit{wikis} are not as good. Similar results are achieved for the advanced features. The results for the vertex balance are shown in Figure~\ref{fig:eval:heatmap-vertex-balance}. The results for the other balancing metrics are omitted here due to space limitations, but are available in our GitHub repository \cite{graphlearner}.

Compared to the replication factor prediction, the prediction results for the vertex balance are less dependent on the graph type, but more on the partitioning algorithm. It is notable that for NE and HEP-100, accuracy is rather low. When investigating the problem in more detail, we found that this is likely related to the unstable behavior of the partitioners themselves: We observed that when NE is executed multiple times on the same graph with the same number of partitions, the vertex balance can heavily fluctuate by a factor of up to 2.1 between different runs. This is due to the random seed vertex selection in NE, which leads to different partitionings in different runs. HEP is, to a lesser extent, affected by the same problem. Different from vertex balance, the replication factor remains stable between different runs of NE and HEP, respectively.   

\begin{table}[t]
  \caption{Quality metrics for \QP {} on test set (average across real-world graphs).} 
  \vspace{-2mm}
  \label{tab:scores-std}
  \begin{center}
  \begin{tabular}{|>{\scriptsize}l|>{\scriptsize}l|>{\scriptsize}l|>{\scriptsize}l|>{\scriptsize}l|>{\scriptsize}l|>{\scriptsize}l|}
     \hline
     \textbf{Target} & \textbf{Model}  & \textbf{Features} & \textbf{MAPE} & \textbf{RMSE}    \\
     \hline\hline
     $RF(P)$ &             XGB &     basic &   0.296 &    1.197 \\
     $RF(P)$ &             XGB &      adv. &   0.288 &    1.238 \\
  $B_{v}(P)$ &             RFR &     basic &   0.152 &    0.921 \\
$B_{\mathit{src}}(P)$ &             RFR &     basic &   0.144 &    0.851 \\
$B_{\mathit{edge}}(P)$ &             RFR &     basic &   0.079 &    1.055 \\
$B_{\mathit{dst}}(P)$ &             RFR &     basic &   0.154 &    1.020 \\
  \hline
\end{tabular}
\end{center}
\vspace{-6mm}
\end{table}
\subsection{Enrichment}
\label{sec:evaluation:enrichment}
We suggest to use real-world graphs to enrich the training set for \QP {} when the model trained with synthetic data shows weaknesses at specific combinations of graph types and partitioners. 

\begin{figure*}[t]
  \centering
  \begin{subfigure}[b]{0.32\linewidth}
      \centering
      \includegraphics[width=\linewidth]{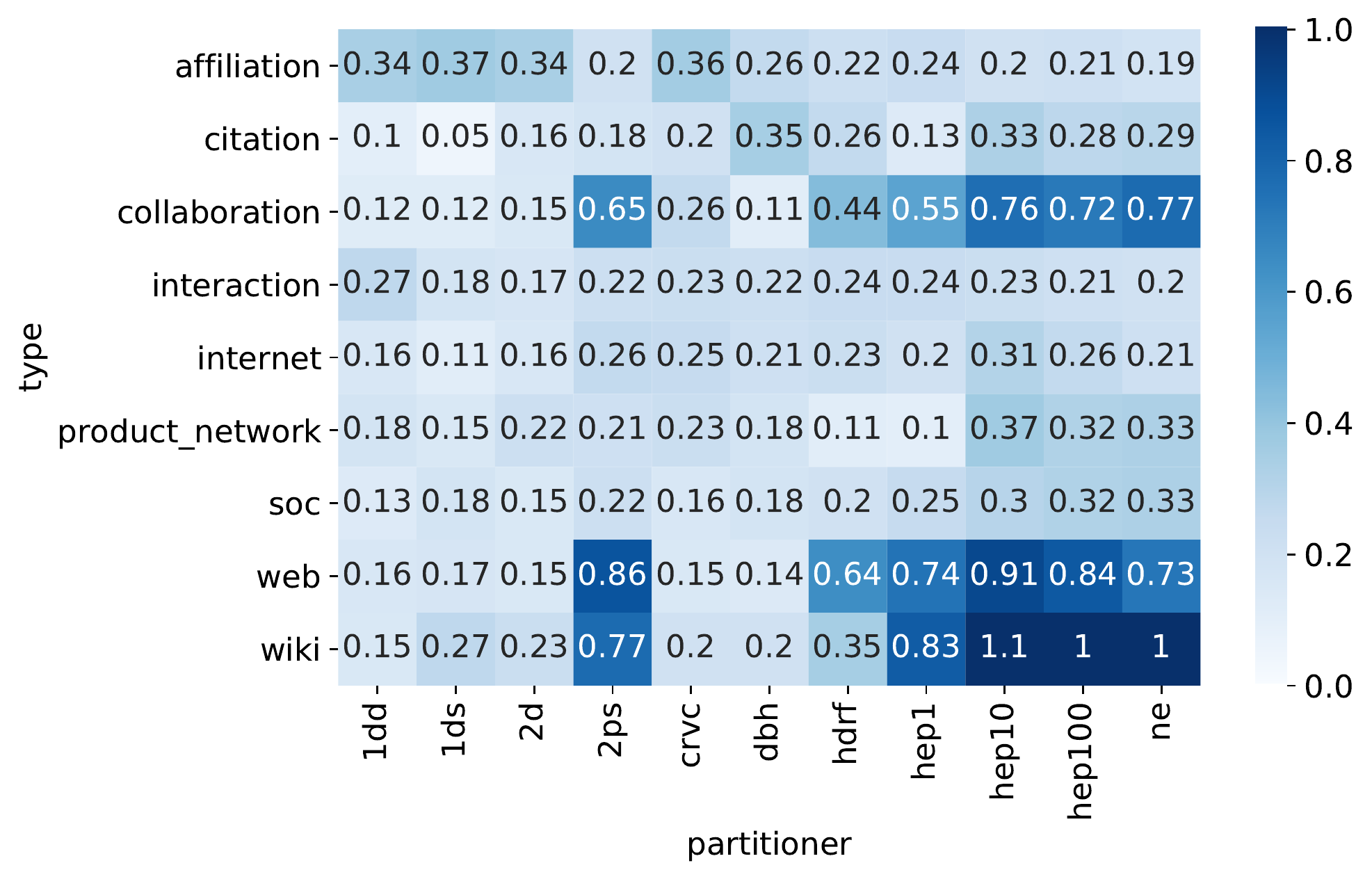}
      \vspace{-4mm}
      \caption{\color{black}Replication factor.\color{black}}
      \label{fig:eval:rfr-basic-0-heatmap}
  \end{subfigure}
  \hfill
  \begin{subfigure}[b]{0.32\linewidth}
      \centering
      \includegraphics[width=\linewidth]{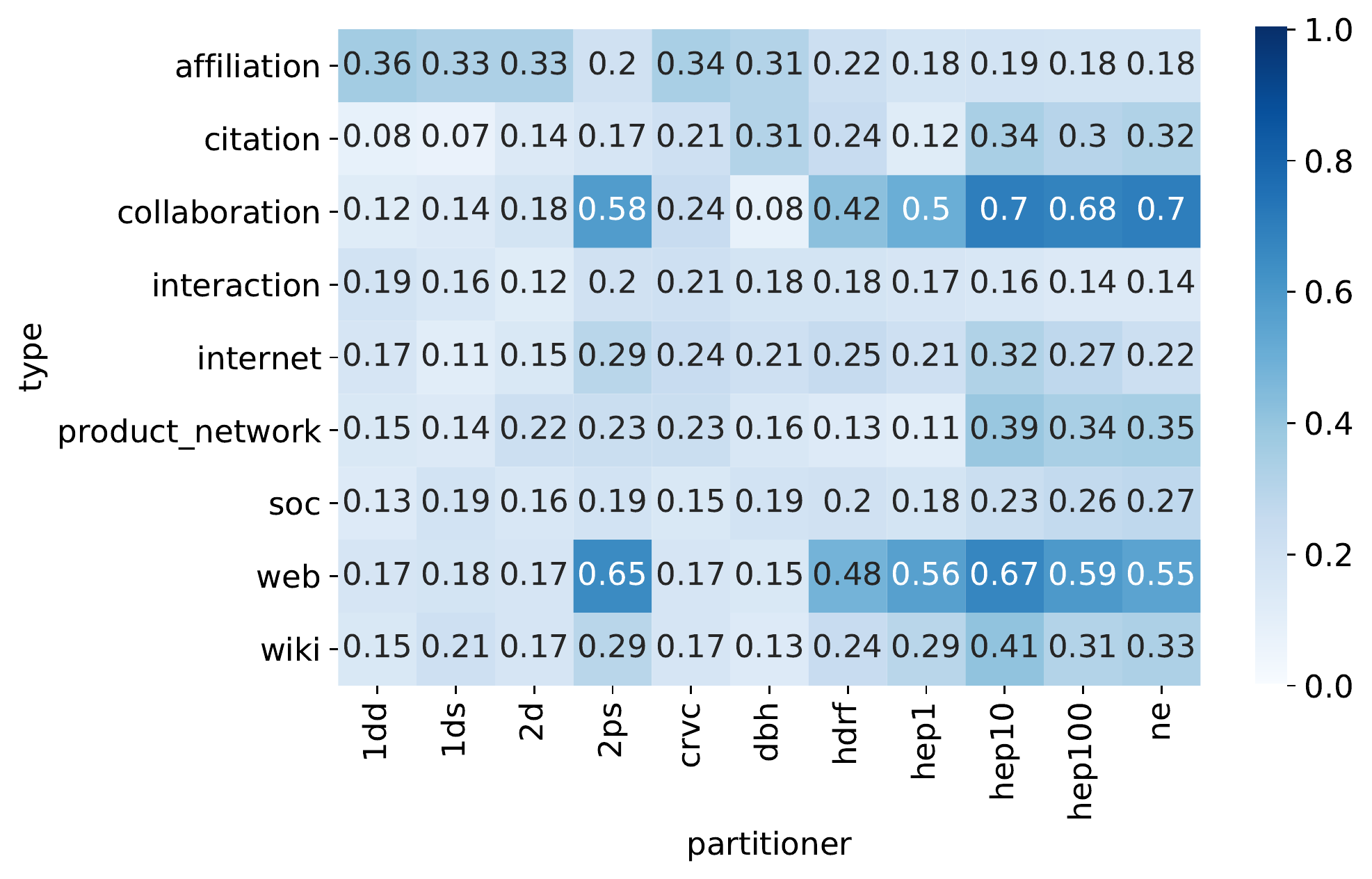}
      \vspace{-4mm}
      \caption{\color{black}Replication factor (with enrichment).\color{black}}
      \label{fig:eval:rfr-basic-100-heatmap}
  \end{subfigure}
   \hfill
   \begin{subfigure}[b]{0.32\linewidth}
       \centering
       \includegraphics[width=\linewidth]{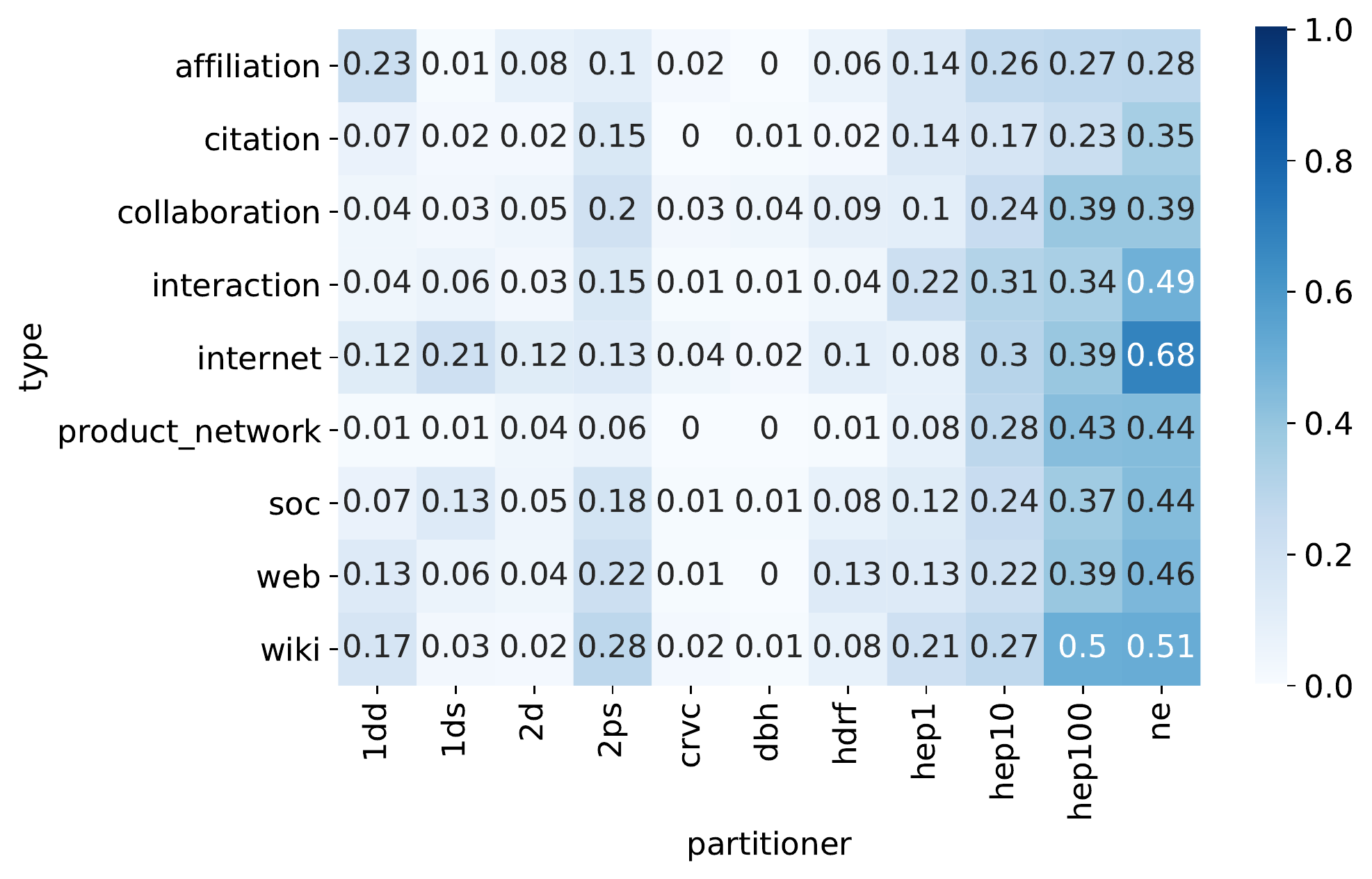}
       \vspace{-4mm}
       \caption{\color{black}Vertex balance.\color{black}}
       \label{fig:eval:heatmap-vertex-balance}
   \end{subfigure}
    \caption{MAPE scores for replication factor and vertex balance predictions.}
    \label{fig:heatmaps}
\end{figure*}

The replication factor predictions for the partitioners 2PS, HDRF, HEP and NE on the graph types \textit{collaboration}, \textit{web} and \textit{wikis} show weaknesses when compared to the other combinations (see Figure~\ref{fig:eval:rfr-basic-0-heatmap}). Hence, we enrich the synthetic training set with up to 96 additional real-world wiki graphs~($G_{\mathit{wiki}}$). In order to investigate how many graphs are needed for the enrichment, we select subsets of 19, 38, 57, 76 and 96 of the wiki graphs~($G_{\mathit{selected}}$) for the enrichment. We use the same test set as in Section~\ref{sec:testset}. Since we randomly select $G_{\mathit{selected}}$ from $G_{\mathit{wiki}}$, we repeat each experiment three times and report the mean quality metrics along with the standard deviation. 

For the balancing metrics and replication factor, we use the RFR models of Table~\ref{tab:scores-std}. XGB only leads to the slightly better MAPE of 0.296 vs. 0.303 when using the basic features, but the training time per enrichment level takes much longer for XGB than for RFR (140 minutes vs. 1 minute).

The results for an enrichment of 96 real-world graphs are shown in Figure~\ref{fig:eval:rfr-basic-100-heatmap} for each combination of partitioning algorithm and graph type. In Figure~\ref{fig:rfr-basic-enrichment-lines}, the influence of the size of the enrichment data set on the quality metrics is shown. The main observations for the replication factor prediction are as follows: 

\begin{figure}[t]
  \centering
  \includegraphics[width=\linewidth]{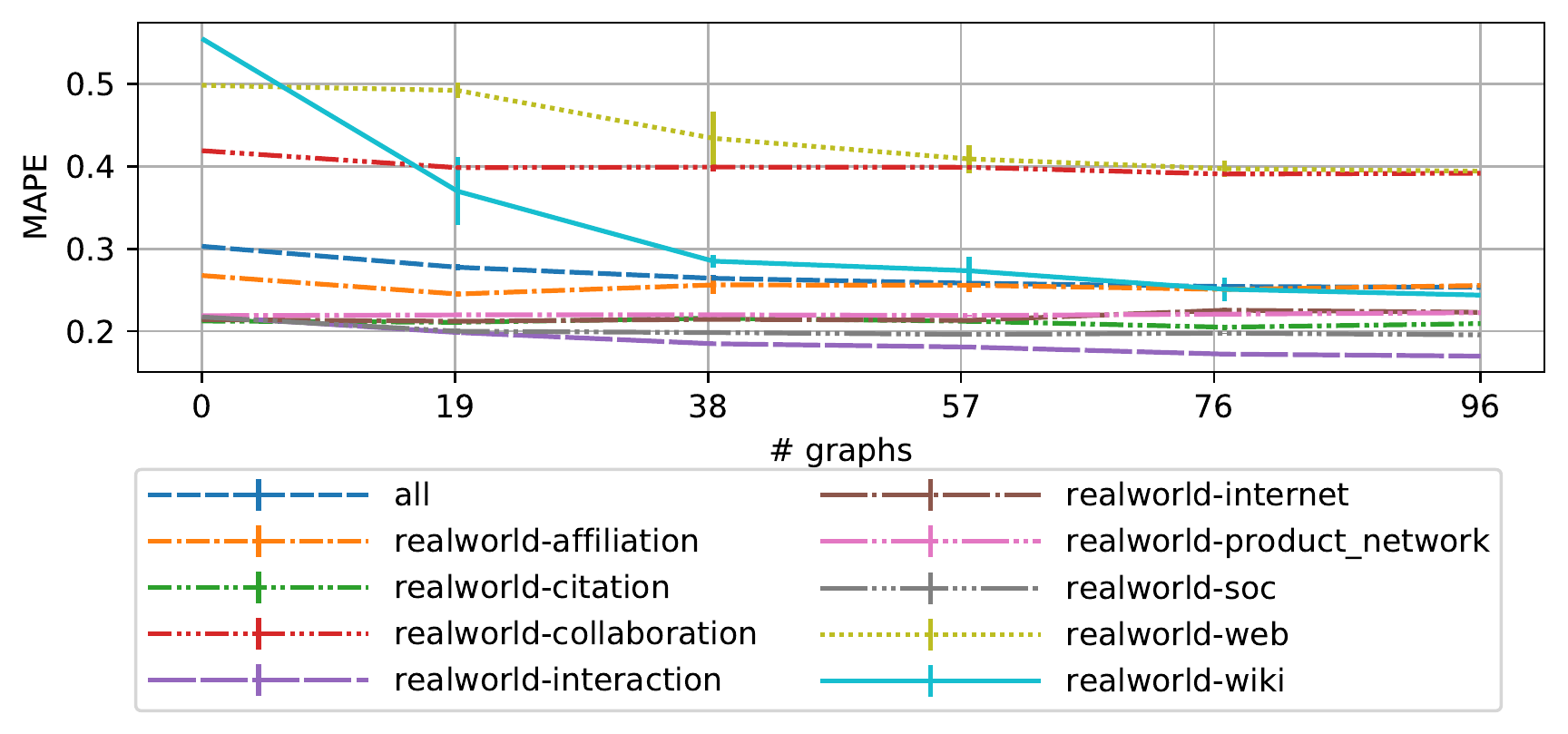}
  \vspace{-5mm}
  \caption{Mean MAPE along with the standard deviation for different graph types with different enrichment levels.}
  \label{fig:rfr-basic-enrichment-lines}
  \vspace{-4mm}
\end{figure}

(1) When enriching with all 96 real-world graphs, MAPE for the wiki graphs decreases from 0.555 to 0.244 and from 0.566 to 0.178 for the basic features and the advanced features, respectively. Therefore, through enrichment, the weakness of the synthetically trained model for the wiki graphs could be reduced.

(2) The advanced features lead to a better MAPE (0.178 vs. 0.242) with enrichment. Therefore, when using enrichment, it is worth to consider the advanced features. A possible explanation for this observation is that through enrichment, wiki graphs with more realistic values for the advanced features are used in the training and therefore, the unseen graphs can be mapped to them better. 

(3) As can be seen in Figure~\ref{fig:rfr-basic-enrichment-lines}, the larger the enrichment, the better the prediction. However, even with an enrichment of only 19 graphs, the prediction quality can be increased a lot (improvement of MAPE from 0.555 to 0.370). In other words, already a small number of real-world graphs in the training data set can improve the prediction for ``weak spots'' of the model.

(4) In addition to the wiki graphs, also the prediction error for other graph types (especially web graphs) was decreased by the enrichment or at least kept stable (see Figure~\ref{fig:rfr-basic-enrichment-lines}). It is expected that the wiki graphs are similar to web graphs and therefore insights gained at the basis of the wiki graphs can be transferred to web graphs.

The prediction performance for the balancing metrics could also be improved for the wiki graphs by enrichment. On average, MAPE slightly decreases from 0.156 to 0.114 for the wiki graphs. However, the overall MAPE for all graph types slightly increases from 0.132 to 0.154. Overall, balancing predictions are already remarkably accurate even without enrichment. Hence, enrichment is less effective in this case. 

\subsection{Feature Importance}
\label{sec:feature_importance} 
In order to understand how important the different features are for the prediction of the partitioning quality metrics, we calculate the feature importance for each feature for the RFR model. A random forest is an ensemble of multiple decision trees. The decision trees split all samples recursively into smaller subsets based on the features. A purity metric is used in order to decide which feature to use for splitting a node. In the machine learning framework we used, the mean squared error is used as the purity metric~\cite{scikit}. The importance of a feature is expressed by how much the purity is increased by splitting a node based on that feature, weighted by how many samples are contained in that node divided by the total number of samples. 

Table~\ref{tab:feature-importance} lists the feature importance of the basic features for all partitioning quality metrics. We make the following observations: First, both the partitioning algorithm and the number of partitions are very important for all partitioning quality metrics, ranging from 0.244 to 0.542 and from 0.177 to 0.477, respectively. Second, the degree distribution is an important feature with 0.165 for the replication factor and a very important feature (0.214 to 0.372) for the balancing metrics. Third, the mean degree is an important feature (0.274) for the replication factor and a less import feature (0.036 to 0.065) for the balancing metrics. The density is for all quality metrics the least important feature (0.007 to 0.034).

\begin{table}[t]
  \caption{Feature importance for quality metrics.} 
  \vspace{-2mm}
  \label{tab:feature-importance}
  \begin{center}
  \begin{tabular}{|>{\scriptsize}l|>{\scriptsize}l|>{\scriptsize}l|>{\scriptsize}l|>{\scriptsize}l|>{\scriptsize}l|}
     \hline
     \textbf{Feature} & $\mathbf{RF}$ & $\mathbf{B_v}$ & $\mathbf{B_{src}}$ & $\mathbf{B_{dst}}$ & $\mathbf{B_{edge}}$    \\
     \hline\hline

     Partitioner &               0.299 &           0.268 &                0.245 &           0.542 &         0.244 \\
     Mean Degree &               0.274 &           0.065 &                0.059 &           0.058 &         0.036 \\
     \#Partitions &               0.256 &           0.271 &                0.293 &           0.177 &         0.472 \\
   Degree Distr. &               0.165 &           0.368 &                0.372 &           0.214 &         0.214 \\
         Density &               0.007 &           0.028 &                0.031 &           0.009 &         0.034 \\
  \hline
\end{tabular}
\end{center}
\vspace{-6mm}
\end{table}
These results seem reasonable. First of all, it is expected that both, the partitioning algorithm and the number of partitions, are important to the machine learning model. Second, it seems plausible that the mean degree is important for the replication factor. We observed for all partitioning algorithms, that the higher the average degree, the higher the replication factor. 
Third, the importance of the degree distribution is also expected. It was already observed that the degree distribution can influence partitioning algorithms~\cite{hdrf, twophasestreaming}. Interestingly, the density is not important. Probably, the mean degree captures similar characteristics and no additional density feature is necessary. 
Therefore, the density could be discarded from the feature set.
\subsection{Automatic Partitioner Selection}
In the following, we evaluate how well \PS{} can automatically select a partitioner based on the predictions of \QP{}, \PaP{} and \PrP{} to minimize the end-to-end time and the graph processing time. We compare \PS {}~(\SPS) against four baseline strategies: (1)~Select the \underline{w}orst partitioner~($S_W$), (2)~select a \underline{r}andom partitioner~($S_R$), select the \underline{o}ptimal partitioner~($S_O$) and (4)~select the partitioner with the \underline{s}mallest \underline{r}eplication \underline{f}actor~(\SSRF). 
\begin{table}[t]
  \caption{Comparison of different partitioner selection strategies: \SPS, $S_O$, \SSRF, $S_R$ and $S_W$ represent the time to which \underline{P}artitioner\underline{S}elector (our approach), the \underline{o}ptimal partitioner, the partitioner with the \underline{s}mallest \underline{r}eplication \underline{f}actor, a \underline{r}andomly selected partitioner and the \underline{w}orst partitioner lead. We differentiate between two selection goals: minimize the end-to-end time (E2E) or the processing time (Pro.).} 
  \label{tab:scores-partitioner-selector}
  \begin{subtable}[t]{1\columnwidth}
    \vspace{0mm}
    \caption{No Enrichment}\label{tab:eval:selection:no-enrichment}
    \vspace{-4mm}
    \centering
  \begin{center}
    \begin{tabular}{|>{\scriptsize}l|>{\scriptsize}l|>{\scriptsize}c|>{\scriptsize}c|>{\scriptsize}c|>{\scriptsize}c||>{\scriptsize}c|}
      \hline
      \multicolumn{1}{|>{\scriptsize}l|}{\multirow{2}{*}{Goal}} & \multicolumn{1}{>{\scriptsize}c|}{\multirow{2}{*}{Algorithm}} &   \multicolumn{4}{>{\scriptsize}c||}{\textbf{$S_{\mathit{PS}}$ in \% of baselines}} & \multicolumn{1}{>{\scriptsize}c|}{\multirow{2}{*}{\shortstack{ \textbf{$S_{SRF}$ in \%} \\ \textbf{of ${S_O}$}}}}\\
      \multicolumn{1}{|>{\scriptsize}c|}{}& \multicolumn{1}{>{\scriptsize}c|}{}  & ${S_O}$ &  $S_{\mathit{SRF}}$ &  ${S_R}$ &  $S_W$ &  \\
      \hline \hline
      E2E &      SSSP &          102 &              69 &          84 &          67 &             152 \\
      E2E &        CC &          103 &              58 &          76 &          57 &             184 \\
      E2E &        PR &          110 &              96 &          96 &          79 &             119 \\
      E2E &   K-Cores &          111 &              76 &          91 &          73 &             152 \\
      E2E &  Synthetic-High &          113 &              98 &          92 &          73 &             119 \\
      E2E &   Synthetic-Low &          117 &              99 &          95 &          76 &             121 \\
      \hline\hline
     Pro. &      SSSP &          106 &              93 &          94 &          80 &             117 \\
     Pro. &        CC &          107 &              92 &          91 &          74 &             121 \\
     Pro. &        PR &          111 &              99 &          96 &          79 &             116 \\
     Pro. &   K-Cores &          113 &              94 &          97 &          80 &             123 \\
     Pro. &  Synthetic-High &         114 &              99 &          92 &          72 &             117 \\
     Pro. &   Synthetic-Low &        119 &              101 &         96 &          76 &             120 \\
      \hline
  \end{tabular}
\end{center}
\end{subtable}
\begin{subtable}[t]{1\columnwidth}
  \vspace{4mm}
  \caption{Enrichment}\label{tab:eval:selection:enrichment}
  \vspace{-4mm}
  \centering
\begin{center}
  \begin{tabular}{|>{\scriptsize}l|>{\scriptsize}l|>{\scriptsize}l|>{\scriptsize}c|>{\scriptsize}c||>{\scriptsize}c|>{\scriptsize}c|>{\scriptsize}c|}
    \hline
    \multicolumn{1}{|>{\scriptsize}l|}{\multirow{2}{*}{Goal}} & \multicolumn{1}{>{\scriptsize}c|}{\multirow{2}{*}{Enrich.}} &   \multicolumn{3}{>{\scriptsize}c||}{\shortstack{\textbf{$S_{\mathit{PS}}$ in \% of baselines}\\ \textbf{(Enwiki-2021)}}} & \multicolumn{3}{>{\scriptsize}c|}{\shortstack{\textbf{$S_{\mathit{PS}}$ in \% of baselines}\\ \textbf{(All Graphs)}}}\\
    \multicolumn{1}{|>{\scriptsize}c|}{}& \multicolumn{1}{>{\scriptsize}c|}{}  & ${S_O}$ &  ${S_R}$ &  $S_W$ & ${S_O}$ &  ${S_R}$ &  $S_W$  \\
    \hline \hline
    E2E &  No &113 & 94&78   &110&89&71\\
    E2E & Yes &107 & 89&74  &112&91&72\\
    \hline \hline
   Pro. & No &111&91&73   &114&96&78\\
   Pro. & Yes &106&88&71&117&98&80\\
    \hline
\end{tabular}
\end{center}
\end{subtable}
\vspace{-6mm}
\end{table}

Our main observations are as follows. In 26.2\% and 35.7\% of the cases, \SPS{} selects the optimal partitioner out of the 11 partitioners to minimize the processing and end-to-end time, respectively. In comparison, manual selection strategies like \SSRF {} and \SR {} fall behind. \SSRF{} selects in 7.1\% and 14.3\% of the cases the best partitioner in terms of processing and end-to-end time, respectively.  
$S_R${} selects only in 9.1\% ($=\frac{1}{11} \cdot 100\%$) of the cases the best partitioner for both optimization goals. 

Further, we analyze the processing time and the end-to-end time to which the selection of \SPS {} lead compared to the four baselines. Table~\ref{tab:eval:selection:no-enrichment}~(see columns 3-6) shows both the processing and end-to-end time of \SPS's selection in percentage of the respective baseline for each graph processing algorithm (lower is better).

\emph{Worst selection $(S_W\! )$:}
\SPS{} leads to a processing time which is on average between 72\% (Synthetic-High) and 80\% (SSSP) of $S_W${} and to an end-to-end time which is on average between 57\% (CC) and 79\% (PR) of $S_W${} (see column $S_W${} in Table~\ref{tab:eval:selection:no-enrichment}). Therefore, \SPS{} leads in all cases to a much better performance.

\emph{Random selection $(S_R\! )$:} 
The end-to-time of \SPS{} is on average between 76\% (CC) to 96\% (PR) of $S_R$ {}and the processing run-time is on average between 91\% (CC) to 97\% (K-cores) of $S_R$ (see column $S_R$ in Table~\ref{tab:eval:selection:no-enrichment}). Therefore, \SPS{} also outperforms random selection. 

\emph{Optimal selection $(S_O\! )$:}
Compared to the optimal selection $S_O$, \SPS {} leads to an end-to-end time which is on average between 102\% (SSSP) and 117\% (Synthetic-Low) of $S_O$ and to a processing time which is on average between 106\% (SSSP) and 119\% (Synthetic-Low) of $S_O$ (see column $S_O$ in Table~\ref{tab:eval:selection:no-enrichment}). Therefore, for the algorithms SSSP, CC and PR, the selection is relatively close to the optimum for both optimization goals, while for K-Cores and the synthetic algorithms there is more room for improvement.

\emph{Smallest replication factor $(S_\mathit{SRF}\! )$:}
\SPS {} leads to an end-to-end time which is on average between 58\% (CC) and 99\% (Synthetic-Low) of \SSRF. The processing time is between 92\% (CC) and 101\% (Synthetic-Low) of \SSRF{} (see column $S_{\mathit{SRF}}$ in Table~\ref{tab:eval:selection:no-enrichment}). Only in \emph{one} case (for the algorithm Synthetic-Low) \SPS {} leads to a graph processing time which is 101\% of \SSRF.

\begin{figure}[t]
  \centering
  \begin{subfigure}[b]{0.49\linewidth}
  \centering
  \includegraphics[width=\linewidth]{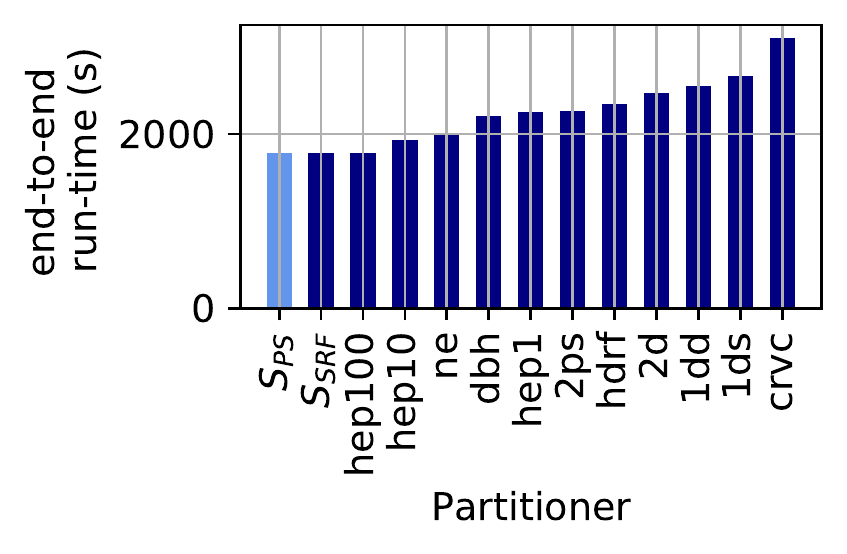}
  \vspace{-6mm}
  \caption{\SPS{} and \SSRF{} select HEP-100.}
  \label{fig:eval:enwiki:synhigh}
  \end{subfigure}
  \hfill
  \begin{subfigure}[b]{0.49\linewidth}
  \centering
  \includegraphics[width=\linewidth]{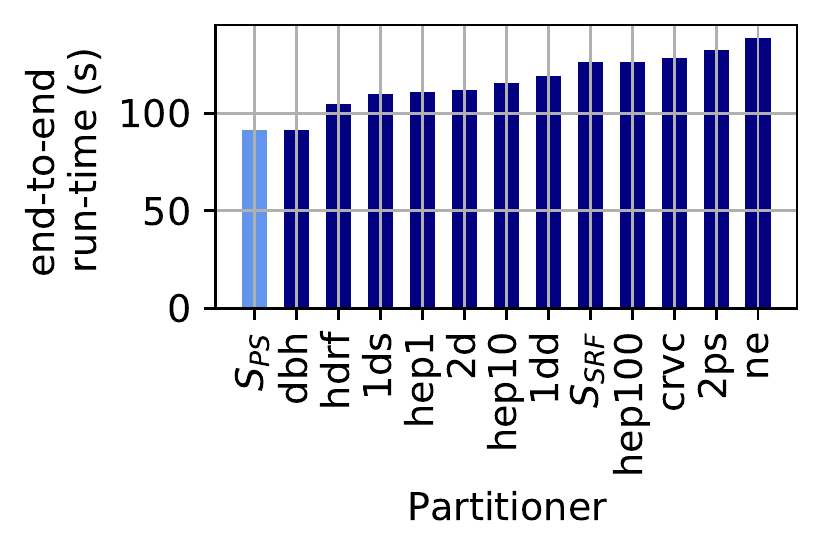}
  \vspace{-6mm}
  \caption{\SPS {} selects DBH. \SSRF {} selects HEP-100.}
  \label{fig:eval:enwiki:cc}
  \end{subfigure}
  \caption{End-to-End time comparison for different partitioners and the selection strategies \SPS{} and \SSRF{} on the graph \emph{Enwiki-2020} for (a) Synthetic-High and (b) CC.}
  \label{fig:eval:enwiki-plot}
  \vspace{-6mm}
  \end{figure}

We further compare \SSRF{} with the optimal strategy $S_O$. In the last column of Table~\ref{tab:eval:selection:no-enrichment}, we report the processing and end-to-end time of $S_{\mathit{SRF}}$'s selection in percentage of $S_O$ for each graph processing algorithm.\color{black} We observe that even if the replication factor of all partitioners would be known, it is not sufficient for partitioner selection. First of all, in addition to the replication factor, the balancing metrics are important. Second, the partitioning time is important and there are cases where it is not worth to invest much time for partitioning to achieve the smallest replication factor, as it can not be amortized by a lower graph processing time. This can be seen in Figure~\ref{fig:eval:enwiki-plot}: HEP-100 achieves the smallest replication factor and minimizes the end-to-end time for the communication-bound algorithm Synthetic-High (see. Table~\ref{fig:eval:enwiki:synhigh}). The partitioning time can be amortized. However, for CC, fast partitioning with DBH leads to the lowest end-to-end time (see Table~\ref{fig:eval:enwiki:cc}). \SPS{} takes into account both partitioning and processing time, and hence, makes a good choice in a wide range of workloads. We further want to point out that \SSRF{} is a rather hypothetical strategy, as the replication factor is usually not known before a graph has been partitioned.

\noindent We also investigated whether through enrichment with the wiki graphs, the performance of \SPS{} for \emph{enwiki-2021} can be improved. We used the same random forest models to predict the partitioning quality with and without enrichment as in the experiments in Section~\ref{sec:evaluation:enrichment}. 
Compared to \SPS{} without enrichment, \SPS{} with enrichment reduces the processing and end-to-end time for \emph{enwiki-2021} on average by 4\% and 5\%, respectively. Evaluated on \emph{all} graphs, the selection of \SPS {} with enrichment increases the processing and end-to-end time on average by 2\% and 3\%, respectively. 
In Table~\ref{tab:eval:selection:enrichment}, we report the average processing and end-to-end time of \SPSS{} selection (with and without enrichment) in percentage of the \emph{respective baseline} evaluated on \emph{enwiki-2021} (see columns 3-5) and evaluated on \emph{all} graphs (see columns 6-8).
Improvements in the target category are larger than the degradation of performance in other categories, so that enrichment pays off if the majority of workloads concern the enriched graph type. \color{black}

\textbf{Summary:} Our experiments showed that \GL {} can provide a \emph{quantitative} prediction of the partitioning run-time and partitioning quality for different types of graph partitioning algorithms. On that basis, \GL {} is able to predict the run-time of graph processing algorithms with different workload characteristics. We achieved a high prediction accuracy by solely training with synthetic graphs which can be improved by enriching the training data with real-world graphs. Based on run-time predictions, we provide automatic graph partitioner selection to either minimize the graph processing or the end-to-end run-time. In both cases, the automatic selection outperforms manual selection heuristics.

\section{Related Work}
\label{sec:relatedwork}

In the past years, experimental studies \cite{survey.1.vldb.2017, survey.3.vldb.2018, survey.2.vldb.2018, survey.4.sigmod.2019, cut.to.fit} have been conducted that provide \emph{qualitative} insights into the characteristics of different graph partitioning algorithms to help selecting one. However, the studies are not sufficient for automatic partitioner selection for a given scenario and do not include an automatic method to incorporate new partitioning and graph processing algorithms. In our work, we provide a \emph{quantitative} prediction of the expected partitioning quality, the partitioning and graph processing time that is easily extensible to include new partitioning and graph processing algorithms.

Lots of research has been conducted in the field of graph generators (e.g., \cite{erdoes, barabasi, watts, rmat, Leskovec.kleinbeg.generator, akoglu2009rtg, bonifati.generator, Chakrabarti.survery}) to generate synthetic graphs with real-world properties. Graph generators are an important component of our approach and we propose to select a suitable graph generator that can generate representative graphs which are expected as a workload and for which the partitioning quality, the partitioning and graph processing run-time should be predicted. In our experiments, we used R-MAT as a domain-agnostic general graph generator \cite{bonifati.generator} which can create graphs with real-world properties \cite{Chakrabarti.survery}. However, further graph generators such as \cite{akoglu2009rtg,generatorsgmwithcommunity,Darwini} can be explored for our approach in the future.

Many graph partitioning algorithms have been developed and can be categorized into (1) streaming algorithms (2) in-memory algorithms and (3) hybrid partitioning algorithms. 
Streaming partitioning algorithms \cite{fennel, stanton,  dbh, powerlyra, grid, graphx, boost, restreaming, adwise,  twophasestreaming,2psl, windowvertexpartitioning, zwolak2022gcnsplit} stream the graph, e.g., as an edge list and assign edges (vertex-cut) or vertices (edge-cut) on the fly to partitions. Streaming algorithms can be stateless or stateful. 
In-memory partitioners \cite{metis, ne, dne, sheep, schulz, spinner, XtraPuLP} load the complete graph into memory to perform partitioning. Therefore, in contrast to streaming partitioning, a complete view of the graph is available for partitioning. This leads in many cases to lower replication factors, but the partitioning run-time and memory overhead may be higher.
Hybrid edge partitioning \cite{hep} is a combination of in-memory and streaming partitioning. One part of the graph is partitioned in-memory and the remaining part in a streaming fashion.
In the evaluation we showed that the partitioning quality and run-time of representative partitioning algorithms of each category could be predicted by our system. We also showed on the example of HEP that our approach can handle partitioners that use a parameter that influences the partitioning quality and the run-time. This could also be applied to other partitioners \cite{CuSP, boost} that have a similar parameter.

Fan et al. \cite{fan2020incrementalization} proposed an approach to incrementalize vertex-cut and edge-cut partitioners. Our approach can be used to select a partitioner, which can then be incrementalized. 
Fan et al. \cite{fan2019dynamic} further studied dynamic scaling for parallel graph processing. In order to scale in or out, the graph often needs to be re-partitioned. Our approach can be applied to predict how the partitioning quality metrics, the partitioning and processing run-time will change. 

Fan et al. \cite{application.driven.partitioning} proposed an application-driven partitioning strategy. They showed how for a graph processing task, a given vertex or edge partitioning can be refined and transformed to a hybrid partitioning. Our approach can be used to support the selection of the initial partitioning which can then be refined and transformed to a hybrid partitioning.

There are several works for using machine learning to optimize data management systems, e.g., to learn index structures \cite{learnedIndexStructures}, cardinality estimation \cite{CardinalityEstimation} or for automatic configuration tuning \cite{AutomaticDatabaseConfigurationTuning}. Our work is among the first approaches to use machine learning to optimize graph processing systems.

\section{Conclusions}
\label{sec:conclusion}
In order to enable distributed graph processing, a graph needs to be partitioned.
However, the plethora of graph partitioning algorithms makes it a challenging task to select a partitioner for a given scenario. 
In this paper, we propose a machine learning-based approach to predict the partitioning quality, the partitioning run-time and the graph processing run-time for different types of partitioning algorithms and graph processing algorithms on unseen graphs. We show that such a quantitative prediction enables automatically selecting the partitioner that minimizes the graph processing or end-to-end time.

In future work, we plan to extend our approach to hypergraph partitioning algorithms \cite{sanders.hypergraphs, socialhashpartitioner, hyperpartitionerKarypis, hypersparse,hype} and systems \cite{hyperxsystem, mesh}. \vspace{1.5mm}

\textit{Acknowledgements:} This work is funded in part by the Deutsche Forschungsgemeinschaft (DFG, German Research Foundation) - 438107855.

\bibliographystyle{IEEEtran}
\bibliography{IEEEabrv,literature}

\end{document}